\documentclass[preprint, 3p]{elsarticle}

\pdfoutput=1
\usepackage{lineno,hyperref}
\modulolinenumbers[5]

\journal{MSSP}

\biboptions{numbers,sort&compress} % to have multiple references as [1-3]
\usepackage{mathptmx}       % selects Times Roman as basic font

\usepackage{makeidx}         % allows index generation
\usepackage{graphicx}        % standard LaTeX graphics tool
\usepackage{multicol}        % used for the two-column index
\usepackage{multirow}

\usepackage{enumitem}
\usepackage{natbib}
\usepackage{float}
\usepackage{transparent}

\usepackage{placeins} % for floatbarrier

\usepackage{amsmath}
\usepackage{amsfonts} 
\usepackage{amssymb}
\usepackage{import}

\usepackage{epsfig}
\usepackage{rotating}
\usepackage{subcaption}
\usepackage{xspace}

\usepackage{capt-of}
\usepackage{bm}
\usepackage{epstopdf}

\usepackage{booktabs,array}

\usepackage{rotating}

\usepackage{pgfplots}
\usepgfplotslibrary{external}% shell-escape beim Aufruf!!!
\usetikzlibrary{plotmarks}
\usetikzlibrary{external}
\pgfplotsset{compat=newest}

\newlength\fwidth

\definecolor{orange}{rgb}{0.87500,0.50781,0.07813}%
\definecolor{purple}{rgb}{0.46094,0.16406,0.51172}%

\definecolor{green1}{rgb}{0.64844,0.85547,0.62500}%
\definecolor{green2}{rgb}{0.19141,0.63672,0.32813}%
\definecolor{green3}{rgb}{0.00000,0.40625,0.21484}%

\definecolor{blue}{rgb}{0.0314,0.3176,0.6118}%

\newcommand{\bs}{\boldsymbol}

\newcommand{\ie}{i.e.\,}
\newcommand{\eg}{e.g.\,}
\newcommand{\cf}{cf.\,}
\newcommand{\etal}{et\,al.\,}
\newcommand{\sref}[1]{Section \ref{sec:#1}}
\newcommand{\aref}[1]{\ref{append:#1}}
\newcommand{\eref}[1]{Eq.\ (\ref{eq:#1})}
\newcommand{\fref}[1]{Fig.\ \ref{fig:#1}}
\newcommand{\tref}[1]{Tab.\ \ref{tab:#1}}
\newcommand{\PNLSS}{PNLSS\xspace}
\newcommand{\BRB}{BRB\xspace}

\newcommand{\RMS}{RMS\xspace}
\newcommand{\ee}{\mathrm{e}}
\newcommand{\ii}{\mathrm{i}}

\newcommand{\real}[1]{\operatorname{Re}\left\lbrace #1 \right\rbrace}
\newcommand{\imag}[1]{\operatorname{Im}\lbrace #1 \rbrace}

\newcommand{\herm}{{}^{\mathrm H}}
\newcommand{\tra}{{}^{\mathrm T}}
\newcommand{\e}[2]{\begin{equation} #1 \label {eq:#2} \end{equation}}
\newcommand{\ea}[2]{
	\begin{eqnarray}
	#1 \label {eq:#2} \end{eqnarray}}

\newcommand{\ForceVec}{\bm{F}}

\newcommand{\timevar}{t}

\newcommand{\ommod}{\tilde\omega}
\newcommand{\Dmod}{\tilde\zeta}

\newcommand{\shpmod}{\tilde{\bm\psi}}
\newcommand{\shpmodlin}{\bm\phi}
\newcommand{\shpmodnorm}{\tilde{\bm\phi}}
\newcommand{\modamp}{q}

\newcommand{\statev}{\bm x}
\newcommand{\inputv}{\bm u}
\newcommand{\outputv}{\bm y}

\newcommand{\statemat}{\mathbf A}
\newcommand{\inputmat}{\mathbf B}
\newcommand{\outputmat}{\mathbf C}
\newcommand{\feedtmat}{\mathbf D}

\newcommand{\noncoeffu}{\mathbf E}
\newcommand{\noncoeffy}{\mathbf F}

\newcommand{\error}{e}

\newcommand{\nonmonu}{\rm \epsilon}
\newcommand{\nonmony}{\rm \chi}

\newcommand{\pnlsserror}{e_\text{rel}}

\makeindex             % used for the subject index
\usepackage{setspace}
\begin{document}

\begin{frontmatter}
\title{Experimental Assessment of Polynomial Nonlinear State-Space and Nonlinear-Mode Models for Near-Resonant Vibrations}

\author[addressILA]{Maren Scheel}
\author[addressIDS]{Gleb Kleyman}
\author[addressImperial]{Ali Tatar}
\author[addressRice]{Matthew R.W. Brake}
\author[addressBosch]{Simon Peter}
\author[addressEindhoven]{Jean-Philippe No\"el}
\author[addressUWM]{Matthew S. Allen}
\author[addressILA]{Malte Krack}

\address[addressILA]{University of Stuttgart, Pfaffenwaldring 6, 70569 Stuttgart, Germany; scheel@ila.uni-stuttgart.de} %,krack@ila.uni-stuttgart.de}
\address[addressIDS]{Leibniz University of Hannover, Appelstrasse 11, 30167 Hannover, Germany}
\address[addressImperial]{Imperial College London, London SW7 2AZ, UK}
\address[addressRice]{William Marsh Rice University, 6100 Main St., 101 Mechanical Engineering Building, Houston, Texas 77005-1827, USA}
\address[addressBosch]{Robert Bosch GmbH, Robert-Bosch-Campus 1, 71272 Renningen, Germany}
\address[addressEindhoven]{Eindhoven University of Technology, Groene Loper 5, 5612 AE Eindhoven, The Netherlands}
\address[addressUWM]{University of Wisconsin-Madison, 1500 Engineering Drive, 525 Engineering Research Building, Madison, WI 53706, USA}

\begin{abstract}
In the present paper, two existing nonlinear system identification methodologies
are used to identify data-driven models. 
The first methodology focuses on identifying the system using steady-state excitations. To accomplish this, a phase-locked loop controller is implemented to acquire periodic oscillations near resonance and construct a nonlinear-mode model. This model is based on amplitude-dependent modal properties, \ie does not require nonlinear basis functions. The second methodology exploits uncontrolled experiments with broadband random inputs to build polynomial nonlinear state-space models using advanced system identification tools. The methods are applied to two experimental test rigs, a magnetic cantilever beam and a free-free beam with a lap joint.
The respective models of both methods and both specimens are then challenged to predict dynamic, near-resonant behavior observed under different sine and sine-sweep excitations.
The vibration prediction of the nonlinear-mode and state-space models clearly highlight the capabilities and limitations of the models. 
The nonlinear-mode model, by design, yields a perfect match at resonance peaks and high accuracy in close vicinity. However, it is limited to well-spaced modes and sinusoidal excitation. The state-space model covers a wider dynamic range, including transient excitations. However, the real-life nonlinearities considered in this study can only be approximated by polynomial basis functions. Consequently, the identified state-space models are found to be highly input-dependent, in particular for sinusoidal excitations where they are found to lead to a low predictive capability.
\end{abstract}


\end{frontmatter}

%%%%%%%%%%%%%%%%%%%%%%%%%%%%%%%%%%%%%%%%%%%%%%%%%%%%%%%%%%%%%%%%%%%%%%%%%%%%%%%%%%%%%%%%%%%%
\section{Introduction}
In the design process of structures, it is crucial to have models that can accurately predict dynamic behavior to ensure safe operation \cite{Ewins2015}. The rising demand for material efficiency and lightweight design has increased the technical significance of \textit{nonlinear} vibrations.
Nonlinear force-displacement relationships occur, for instance, due to large displacements or contact interactions in joints.
Linear structural design can be outperformed by mastering nonlinear effects, and nonlinear behavior can even be exploited, for example in nonlinear absorbers  \cite{Vakakis2008}. 
However, simulating nonlinearities, in joints for example, is not a straightforward process and remains a modeling challenge \cite{Popp2003,Brake2018}.
In such cases, directly identifying nonlinear models based on experimentally obtained data can be beneficial. Other use cases for identifying models experimentally include model updating or structural health monitoring.

Nonlinear system identification is a vast and active field of research. Its application to structural dynamics is covered in several review articles \cite{Kerschen2006, Noel2017b}. There is no nonlinear system identification method that is well suited for \textit{all} applications. The methods differ significantly by the dynamic regime of interest, the type of model that is obtained, and the required a-priori knowledge of the nonlinearity as well as the excitation signal and mechanism to obtain the training data for identifying the model.
It is out of scope of this introduction to review all existing nonlinear system identification methods. Instead, we focus on what we deem the most mature classes of methods suited for both stiffness and damping nonlinearities, caused for example by joints.

One very popular experimental approach in both linear and nonlinear structural dynamics is frequency response testing, \eg with stepped sine measurements or sine sweeps. To measure multivalued frequency responses, exhibited by many nonlinear systems, the system has to be excited with both increasing and decreasing frequencies (\eg \cite{Setio1992}) or excited using a controller \cite{Peter2018,Barton2012}.
The concept of frequency response functions (FRFs) can be generalized, for example with higher-order FRFs \cite{Chatterjee2003a,Chatterjee2003b,Lee1997,Lin2018} or nonlinear output FRFs (see \eg \cite{Peng2008}). Both concepts are based on Fourier transforms of Volterra series and have been utilized for parametric identification using measured higher-order or nonlinear output FRFs. To this end, however, the functional form of the nonlinearity must be known. Moreover, the measurement of these generalized FRFs is elaborate.
Also, this approach was so far mostly applied only numerically and on systems with only a single \cite{Chatterjee2003b,Lee1997} or few degrees of freedom \cite{Lin2018,Peng2008} and experimental applications are scarce (\eg \cite{Chatterjee2003a}).

One approach for identifying nonlinear models of systems with many degrees of freedom and nonlinearities is nonlinear subspace identification \cite{Marchesiello2008, Noel2013,Noel2014}. The nonlinearities are interpreted as output feedback to a linear structure, where the nonlinearity is described using basis functions of the output. These functions must be chosen a-priori, for instance based on a previous characterization step. 
A more general model is a polynomial nonlinear state-space (\PNLSS) system, first introduced in \cite{Paduart2010}. The \PNLSS model is a linear, time-invariant state-space model, extended with multivariate polynomial nonlinearities in states and inputs. In principle, multivariate polynomials are able to treat a wide class of nonlinearities \cite{Paduart2010}, such as hysteretic nonlinearities \cite{Noel2017a}.
Generic input signals can be used as training data, among them periodic signals, sine sweeps, random noise, or application specific signals (see \eg \cite{Paduart2010,Cooper2018b, Decuyper2018, Widanage2011}). The resulting model quality depends on the training data \cite{Widanage2011}.

Nonlinear modal models are defined by modal properties, that is, by the modal frequency, damping ratio, and deflection shape, all of which depend on the vibration level (in the nonlinear case). It is not required to a-priori choose the mathematical form of the nonlinearity.
The available nonlinear modal analysis methods differ by the considered definition of nonlinear modes, the excitation strategy (\eg impact testing or force appropriation) and the method to extract the modal properties from the response data.
Using impact testing, one can track the instantaneous frequency and damping ratio, \eg using short-time Fourier transform. However, mode coupling due to nonlinear forces can distort the extracted modal properties \cite{Kuether2016}. To de-convolute the modes from the response, empirical mode decomposition and slow-flow analysis can be useful \cite{Eriten2013}.
With force-controlled FRFs at different amplitudes, amplitude-dependent modal parameters can estimated using linear methods  (\eg \cite{Lin1993}). This method, however, fails for multivalued FRFs. One remedy is to control the response level, which leads to linearized FRFs \cite{Link2011}. In both cases, however, measuring FRFs at many forcing levels is a time-consuming procedure, and this raises the risk of damage during testing.
Therefore, approaches have been suggested to obtain the amplitude-dependent modal properties directly, without the need of measuring FRFs, in particular by extending the idea of phase resonance testing to the nonlinear case. 
Peeters \etal \cite{Peeters2011a} were the first to propose a method to isolate a nonlinear normal mode with force appropriation based on a phase resonance condition. Once the mode is isolated, the excitation is removed and modal properties are extracted from the free decay. An inherent drawback of this method is that only a limited signal length (number of pseudo-periods) can be recorded, determined by the damping. This limits the resolution of the identified modal properties and makes the results sensitive to noise. Moreover, switching off the excitation without inducing perturbations by the excitation mechanism (\eg in form of an impulse) is challenging in practice.
Alternatively, phase resonance approaches with steady-state measurements can be achieved using stabilizing phase controllers \cite{Peter2017, Denis2018} or control-based continuation \cite{Renson2016b}.
The aforementioned modal approaches rely on so-called nonlinear normal modes, which are defined for the underlying undamped system. Damping is either assumed to be small and negligible (in free decay measurements) or considered parasitic and compensated by the excitation force (in steady-state measurements).
For many cases, however, damping is considerable, nonlinear and cannot be neglected. In this case, the extended periodic motion concept introduced by Krack \cite{Krack2015}, which generalizes the concept of nonlinear modes to systems with nonlinear damping, is more appropriate. In \cite{Scheel2018, Schwarz2019}, phase resonance testing with controlled excitation was used to isolate and identify nonlinear modes in accordance with the extended periodic motion concept. Assuming that the vibration energy is confined in only one nonlinear mode, the nonlinear system can be well modeled as a single-degree-of-freedom oscillator, the parameters of which are the nonlinear modal properties. This simplification is justified, in particular, near isolated resonances (under essentially harmonic excitation).

As mentioned above, no nonlinear system identification method is suited for all purposes, meaning each method has distinct limitations. However, these limitations are not always clearly stated and therefore not obvious to prospective users.
Well-founded knowledge of both capabilities and limitations is required to properly select and successfully apply nonlinear system identification. With this paper, we contribute to this knowledge by assessing two representative methods from two fundamentally different clases of system identification methods. Comparing these methods directly for the same task and specimens, we pinpoint each method's strengths and weaknesses, focusing on the models' ability to predict vibrations near resonances.
The two chosen approaches are (a) the identification of a nonlinear-mode model as recently proposed in \cite{Scheel2018} and (b) identification of a \PNLSS model as suggested in \cite{Noel2017a} as example for an already established black-box approach. They rely on fundamentally different input signals: The nonlinear-mode model is identified using controlled, steady-state oscillations, whereas the \PNLSS model is, in this study, trained on broadband multisine excitation.
As indicator for the identified models' quality, we asses their capability to predict vibrations under sinusoidal excitation in the frequency range around an isolated modal frequency. Both steady-state conditions and a transient case with time-dependent frequency are analyzed. The models' quality and dependence on the training data are discussed and evaluated based on these predictions.
The methods are compared for two specimens: a thin beam with magnets at the tip exhibiting a hardening stiffness nonlinearity \cite{Kleyman2017} and a jointed beam, known as the Brake-Reuss beam (\BRB) \cite{Brake2019}. Due to friction in the joints, both modal frequency and damping ratio depend on the vibration level.
 
The paper is structured as follows. In Section \ref{sec:methods}, the applied identification methods are briefly reviewed. The setup and results for the beam with magnets and the \BRB are shown in Section \ref{sec:MagBeam} and \ref{sec:BRB}, respectively. The paper is concluded in Section \ref{sec:conclusion}.

%%%%%%%%%%%%%%%%%%%%%%%%%%%%%%%%%%%%%%%%%%%%%%%%%%%%%%%%%%%%%%%%%%%%%%%%%%%%%%%%%%%%%%%%%%%%
\section{Methods for Identifying Nonlinear Models} \label{sec:methods}

\subsection{\PNLSS model}\label{sec:pnlss} 

This section is a brief recap of the method introduced in \cite{Paduart2010}.
In general, a \PNLSS model is a time-discrete multiple-input-multiple-output state-space model which is composed of a linear state-space model (coefficient matrices $\statemat, \inputmat, \outputmat, \feedtmat$) plus polynomial nonlinear terms (with coefficient matrices $\noncoeffu$ and $\noncoeffy$). 
The inputs exist as $\inputv \in \mathbb{R}^q$, \eg excitation forces applied via a shaker. Further, the vibration response of the system is expressed in terms of the outputs $\outputv \in \mathbb{R}^{l}$, \eg measured with accelerometers. The \PNLSS model is then
\ea{\left\lbrace 
	\begin{aligned} 
		\statev (\timevar_{k + 1}) &= \statemat \statev (\timevar_k) + \inputmat \inputv (\timevar_k) + \noncoeffu \nonmonu (\statev(\timevar_k), \inputv(\timevar_k))\\
		\outputv (\timevar_k) &= \outputmat \statev (\timevar_k) + \feedtmat \inputv (\timevar_k) + \noncoeffy \nonmony (\statev(\timevar_k), \inputv(\timevar_k)).
	\end{aligned} \right.
}{statespace}
Here, $\timevar_k$ are discrete equidistant time levels, $\statev \in \mathbb{R}^{s}$ is the state vector, and $\statemat \in \mathbb{R}^{s \times s}$, $\inputmat \in \mathbb{R}^{s \times q}$, $\outputmat \in \mathbb{R}^{l \times s}$, $\feedtmat \in \mathbb{R}^{l \times q}$ are the linear state, input, output and feedthrough matrices.

The nonlinear basis functions are polynomials both in states and inputs. Multivariate polynomials are straightforward to compute and a wide class of nonlinearities can be modeled by polynomials \cite{Paduart2010}.
The monomial combinations of degree 2 up to degree $d$ are listed in $\nonmonu \in \mathbb{R}^{n_e}$ and $\nonmony \in \mathbb{R}^{n_f}$ with their coefficients gathered in $\noncoeffu \in \mathbb{R}^{s \times n_e}$ and $\noncoeffy \in \mathbb{R}^{l \times n_f}$. In this study, the nonlinear basis functions $\nonmonu$ and $\nonmony$ are the same in both equations of \eref{statespace} and $n_e = n_f$. In general, however, one can use different sets or different polynomial degrees.

To predict the response, given a known input force $\inputv(\timevar_k)$, the \PNLSS model is evaluated successively from given initial values for the time span of interest.

\subsubsection{Excitation with Random Input}\label{sec:pnlss_excitation} 

In this work, the excitation and identification procedure described in \cite{Noel2017a} is followed using the \PNLSS MATLAB toolbox \cite{Tiels2016}. 
Multisine signals that have a discrete frequency spectrum with prescribed magnitudes, random phases, and a constant magnitude over a certain frequency band and zero magnitude otherwise are used for the \PNLSS identification. The phases for each frequency line are drawn from a uniform distribution. The crucial parameters that the user has to specify are the frequency band and the magnitude. Multisines are popular as appropriate signal generators are available in many commercial testing packages.

The actual excitation signal consists of several repetitions (called blocks) of one multisine realization, \ie the signal generated from the same draw of the random phase distribution. This way, transients decay during the first few blocks, and are discarded for the identification. 
To gather training data, several experiments are conducted, each experiment with a different realization, \ie the same amplitude spectra but different phase realizations in accordance with the uniform distribution.
The excitation amplitude spectrum must be chosen carefully such that the structure's nonlinearity is excited in the range of interest. Theoretically, if the nonlinear basis functions can replicate the dynamics and the model order is sufficiently high, the identified model parameters are independent of the excitation level. This, however, will generally not hold. Friction, for example, cannot be modeled accurately with polynomials, which causes an inevitable model error. Therefore, the identified model parameters depend on the input, particularly on the level of the input.

\subsubsection{Identification}

Identifying a \PNLSS model, the parameters $(\statemat, \inputmat, \outputmat, \feedtmat, \noncoeffu, \noncoeffy)$ must be fitted to measured data. To this end, a nonlinear optimization problem is solved with the well-known Levenberg-Marquardt algorithm. The cost function is the squared difference between the simulated $\outputv_{\text{sim}}$ and the measured output $\outputv_{\text{ref}}$ in time domain,
\ea{ \begin{aligned} & \underset{(\statemat, \inputmat, \outputmat, \feedtmat, \noncoeffu, \noncoeffy)}{\text{minimize}}  \quad \sum_{k} 	\left\lVert \bm \error(\timevar_k) \right\rVert^2 \\
		&\text{with }  \bm \error(\timevar_k) = \outputv_{\text{ref}}(\timevar_k) - \outputv_{\text{sim}}(\timevar_k).
\end{aligned}}{opt_nonlin}
It is crucial to have a good initial guess for the optimization. Therefore, a linear state-space model with $\noncoeffu = \bm 0 = \noncoeffy $ is fitted using frequency-domain subspace identification \cite{McKelvey1996,Pintelon2012} and further improved by nonlinear optimization in the frequency domain. Then, this linear model is extended with $\noncoeffu, \noncoeffy \neq \bm 0$ and optimized in accordance with \eref{opt_nonlin}.  

The goal of the identification is, of course, to minimize the deviation with respect to each output quantity at each time level. The optimization algorithm, however, only sees the scalar integral measure in \eref{opt_nonlin} as cost function. If the training data is composed of data points mainly in the low-level regime with only few data points in the high-level range, the latter is under-represented. Then, the numerical optimization algorithm minimizes the error in the low-level (almost linear) range, with less weight on the high-level regime, even though it is typically more relevant to the environment of interest.

\subsection{Nonlinear-Mode Model} 

The modal identification approach in this work is taken from \cite{Krack2015,Scheel2018}.
In \cite{Krack2015}, the nonlinear mode is defined as a periodic motion with the modal frequency $\ommod$, modal damping ratio $\Dmod$, and the Fourier coefficients of the periodic motion for all harmonics, $\lbrace \shpmod_0\, \shpmod_1\, \shpmod_2\, \shpmod_3 ... \rbrace$, where $\shpmod_i$ is the $i$-th harmonic of the deflection shape. The properties depend on the vibration level, indicated by the tilde $\tilde{(\bullet)}$. 

The identified nonlinear-mode model is a single-degree-of-freedom oscillator that replicates the dynamics under the assumption that the system responds only in a single nonlinear mode in either the autonomous case or under mono-frequency excitation. It is common to superimpose the contribution of the remaining modes in a linear way, in order to improve the accuracy further away from resonance \cite{Peter2018}. This was, however, not done here as it had no significant effect in the narrow frequency bands considered in this study.
Assuming further that the system responds essentially in a periodic way with only slowly varying properties, the response of the system is given by the differential equations \cite{Krack2014a}
\e{\begin{bmatrix} \dot{\modamp} \\ \dot{\theta} \end{bmatrix} = 
	\dfrac{1}{2 \Omega} \begin{bmatrix} -2 \Dmod \ommod \Omega \modamp + \imag{ \shpmodnorm\herm_1 \ForceVec_{1,\rm exc} \ee^{-\ii \theta } } \\ \ommod^2 - \Omega^2 -  \frac{1}{\modamp} \real{ \shpmodnorm\herm_1 \ForceVec_{1,\rm exc} \ee^{-\ii \theta } }
\end{bmatrix},}{slowflow} 
where $\modamp$ and $\theta$ are the magnitude and the phase of the modal amplitude, and $\ForceVec_{1,\rm exc}$ and $\Omega$ are the fundamental harmonic magnitude and frequency of the excitation force. The Hermitian transpose ${(\bullet)}\herm$ is used here, and $\shpmodnorm_1$ is the fundamental harmonic of the mass-normalized deflection shape.
The assumption of slow variation of amplitude, phase lag and frequency compared to the oscillation is an inherent limitation of the method and thus of the nonlinear-model model.

\subsubsection{Excitation with Controlled Input} \label{sec:modal_excitation}

To experimentally isolate the mode under investigation, the periodic motion is ensured with an appropriate external excitation.
It was shown in \cite{Scheel2018} that, in many cases, a nonlinear mode can be well-isolated with a single-point excitation and ensuring only local phase resonance with regard to the fundamental harmonic.
This is ensured using a phase controller, called phase-locked loop (PLL), which adapts the excitation frequency until the desired phase lag is reached. The controller is shown schematically in \aref{setup_parameters}. 
The amplitude of the fundamental harmonic of the excitation is set by the user. Therefore, the amplitude during measurement is changed to follow the backbone curve of the system.

\subsubsection{Identification}

The modal frequency $\ommod$ as well as the deflection shape $\lbrace \shpmod_0\, \shpmod_1\, \shpmod_2\, \shpmod_3\, ... \rbrace$ are extracted directly from the measurements.
For a periodic motion, the dissipated power must be equal to the excitation power, averaged over one vibration cycle.
Considering only the fundamental harmonic component of the active power $P_1$ provided by the external force, the modal damping ratio is defined as \cite{Scheel2018} 
\e{\Dmod = \dfrac{P_1}{\ommod^3 \modamp^2}.}{dmod}
The modal amplitude is the scaling factor between the fundamental harmonic of the mass-normalized and the unscaled deflection shape, $\shpmod_1 = \modamp \shpmodnorm_1$ \cite{Scheel2018}.
To compute this factor, linear mass-normalized mode shapes $\bm\Phi$ are required and obtained with a second measurement, a standard linear experimental modal analysis (EMA) at low excitation level. With these, the modal amplitude is determined as \cite{Scheel2018}
\e{\modamp^2 = \shpmod_1\herm \left( \bs\Phi\tra \right)^+ \left(\bs\Phi\right)^+ \shpmod_1.}{modalampl}
Here, ${(\bullet)}\tra$ is the transpose and ${(\bullet)}^+$ the pseudoinverse.

\subsection{Expected strengths and weaknesses}

In this section, the key strengths and weaknesses of the two approaches are briefly summarized:
The nonlinear-mode model does not make any assumption on the functional form of the nonlinearity. In the experiment, controlled excitation is required, which can take time and effort (on-line cost). There is a risk that the controller is invasive, \ie modifies the dynamics compared to the underlying uncontrolled experiment, and, thus, that the identified model is sensitive to the control parameters.
The nonlinear-mode model is valid in the amplitude range that was identified. Outside of this range, extrapolation errors are expected. 
Moreover, the model is specifically designed to describe behavior near isolated modes, which is an inherent limitation. It is only applicable to model the response under essentially mono-harmonic excitation, potentially with slowly varying parameters, \eg excitation frequency and level. By design, a perfect match of the model with phase-resonant points of the frequency response is, theoretically, expected. However, a deviation can occur in practice due to repetition-variability or interpolation error as shown later.

The \PNLSS model is a parametric model of the nonlinearity and assumes the functional form as multi-variate polynomials.
Arbitrary excitation signals are possible with the benefit that no control is needed. Band-limited random signals are attractive because multiple modes can be excited at once and subspace identification of linear models can easily be applied, which serves as an initial guess for the nonlinear identification.
The benefit of the state-space model is that one can simulate the response to any excitation signal and also determine transients.
Theoretically, the validity range covers quite generic dynamic regimes, \ie the regime is not a-priori limited to isolated modal frequencies and harmonic, near-resonant excitations. 
However, the valid amplitude range is practically limited due to the modeling error and the finite truncation order - the computational effort grows exponentially with the polynomial truncation order. Additionally, extrapolation should be avoided (\ie the values of the state variables should not exceed the values excited by the training data). Moreover, the regime of interest should be represented well in the training data. The valid frequency range is defined through the design of the multisine input.
Identified \PNLSS models have several error sources: (a) truncation leads to model error inevitable in reality, even for nonlinearities with convergent Taylor series expansion; (b) the time discretization must be fine enough to represent the dynamics and the excitation input accurately; (c) the numerical optimization algorithm is likely to end up in a local minimum and so benefits from suitable initial guesses to find a global minimum.

In summary, only \PNLSS models understand not only harmonic input and can (theoretically) represent several modes at the same time. Also, the excitation does not need to be controlled, which makes it easier to practically realize without the risk of modifying the dynamics. The nonlinear-mode model seems ideal for behavior near isolated resonances, which is often of primary interest for vibration problems.
Further, it is expected that the \PNLSS models can predict transient sine sweeps with good accuracy. But how well do \PNLSS models predict harmonic vibrations close to resonance? And how well does the nonlinear-mode model predict transient sine sweeps? In this work, we answer these questions and \textit{challenge} the models to assess the capability and especially the limitations of both approaches in predicting vibrations.
On the one side, nonlinear-mode models are identified for a defined range of harmonic excitation amplitudes. Then, the models are used to simulate (not-necessarily slow) sine sweeps, which are often of technical relevance, \eg for resonance crossing during acceleration of rotating machinery. This way, the boundary of the nonlinear-mode model is tested, since this model type relies on slow variation of parameters.
On the other side, \PNLSS models are identified even though the nonlinearities in the specimens are probably \textit{not} well-described by low-order polynomials. Moreover, we train the model with an excitation spectrum (multisines), which leads to a minimal experimental effort. The training data, however, is much different from the operating regime of interest (harmonic, \ie single-sine).
In particular, one could expect that for the considered excitation spectrum, the high-level near-resonant behavior might not be represented well in the training data. Thus, the less relevant off-resonant response might inherently receive unreasonably high weight, so the model cannot be expected to be highly accurate near resonance.

%%%%%%%%%%%%%%%%%%%%%%%%%%%%%%%%%%%%%%%%%%%%%%%%%%%%%%%%%%%%%%%%%%%%%%%%%%%%%%%%%%%%%%%%%%%%
\section{Application to a Cantilevered Beam with Repelling Magnets} \label{sec:MagBeam}

The first specimen is a cantilever aluminum beam with the dimensions given in \fref{MB_setup}.
Two ferrite magnets are attached near the free end, close to the node of the second bending mode. The magnets have a cylindrical shape and a residual induction of about 1.1 T.
Opposite of these magnets, at a distance of $d_0 = 11 \text{mm}$, two magnets with same specifications are attached to a frame, causing a nonlinear magnetic repulsive force $f$ acting between the magnet pairs.

\begin{figure} % experimental setup MagBeam
	\centering
	\begin{subfigure}{0.49\textwidth}
		\centering
		\def\svgwidth{0.9\textwidth}
		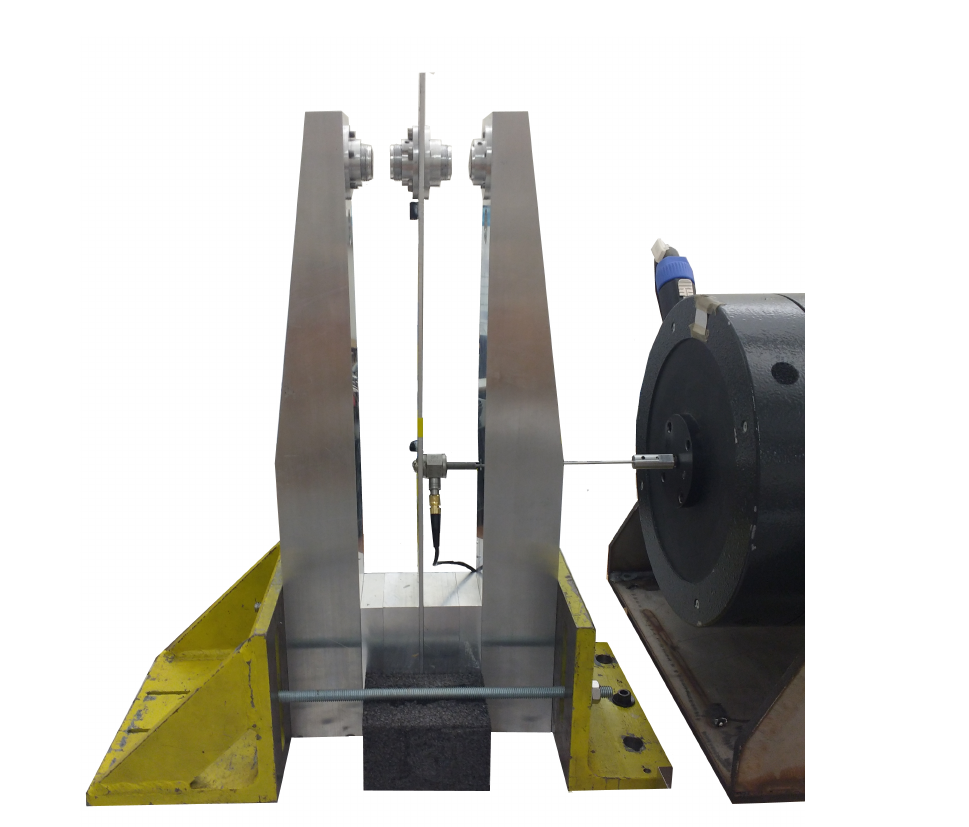\caption{}
	\end{subfigure}
	\begin{subfigure}{0.49\textwidth}
		\centering
		\def\svgwidth{\textwidth}
		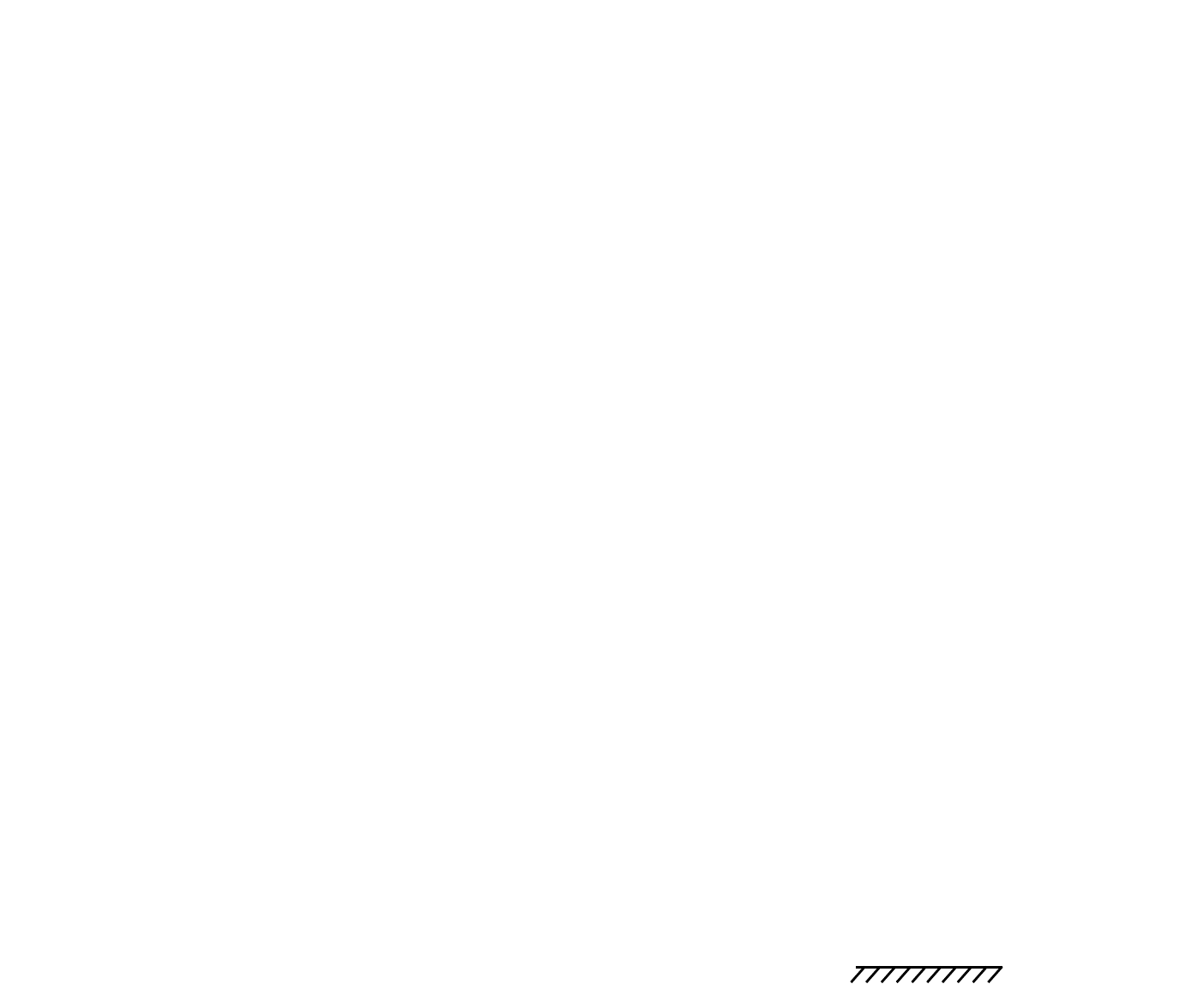\caption{}
	\end{subfigure}
	\caption{Picture (a) and scheme (b) of the aluminum beam with magnets and the experimental setup.}
	\label{fig:MB_setup}
\end{figure}

\subsection{Experimental Setup and Testing Procedure}

In all experiments of this specimen, the excitation was applied with a Br\"uel \& Kj\ae r Vibration Exciter Type 4808 and a 100 mm long steel rod stinger, 2 mm in diameter. To measure the excitation force, a PCB 208C01 force sensor was screwed to the beam. Two uniaxial accelerometers (PCB 352A24) were glued to the beam, one close to the excitation location and one close to the magnets. All results in the following will be shown for the sensor close to the magnets. 

The study's measurement campaign is composed of five different types of measurements: one for a linear modal analysis, one for the nonlinear modal analysis, one for \PNLSS identification, and two types as reference measurements for subsequent vibration prediction. The different types of measurements are listed in \tref{MagBeam_measurement}, including the measurement system and the order in which they were conducted. 

\begin{table} % table with validation errors
	\centering
	\begin{tabular}{ l l l l }
		\toprule
		& Purpose & Excitation &  Data acquisition platform \\
		\midrule[0.8pt]
		1 & reference & (stepped) sine (PLL) & dSPACE MicroLabBox\\
		2 & linear modal analysis & multisine & LMS Test.Lab / Siemens SCADAS Mobile\\
		3 & nonlinear modal analysis & force appropriation (PLL) & dSPACE MicroLabBox \\
		4 & \PNLSS identification & multisine & LMS Test.Lab / Siemens SCADAS Mobile \\
		5 & reference & sine sweeps (uncontrolled) & LMS Test.Lab / Siemens SCADAS Mobile\\
		\bottomrule
	\end{tabular}
	\caption{Measurement campaign for both specimens. The measurements are listed in the order they were conducted.} \label{tab:MagBeam_measurement}
\end{table}

A linear experimental modal analysis (EMA) was performed with pseudo-random multisine signal from 10 to 500 Hz. The first two bending modes were identified in this range (see \tref{MagBeam_linearmodal}). Both modes are lightly damped and well separated. Due to the location of the magnets, the second mode is only slightly affected by the nonlinear force \cite{Kleyman2017}. Thus, the following experimental study focuses on the beam's first bending mode. 

\begin{table} % table with linear modal properties
	\centering
	\begin{tabular}{ l l l }
		\toprule
		& first bending mode & second bending mode \\
		\midrule[0.8pt]
		linear modal frequency $\omega$ & 24.37 Hz & 134.06 Hz \\
		linear modal damping ratio $\zeta$ & 0.43 \% & 0.07 \% \\
		\bottomrule
	\end{tabular}
	\caption{Linear modal frequency and damping ratio for the first two bending modes, identified with EMA at low level.}\label{tab:MagBeam_linearmodal}
\end{table}

\subsection{Model Identification}

\subsubsection{\PNLSS Models}\label{sec:BRB_PNLSS_models}

The nonlinear magnetic repulsive force $f$ acting between the magnet pairs is \cite{Yung1998}
\e{f(z) = -\frac{K_0}{(d_0-z)^4}.}{force_magnetic}
The force depends on a constant $K_0$ associated with the magnetic and geometrical properties and on the deflection $z$ of the beam at the height of the magnets in horizontal direction (see \fref{MB_setup}). This force-deflection-relationship is only valid if the distance between the magnets is larger than the radius of the magnets. Further, it is assumed that the magnetic dipoles are axially aligned with each other and only perform purely translational motion in the $z$-direction, which is a valid assumption for small deflections.
The nonlinear force has an infinite-order Taylor series approximation. Therefore, a truncation error is inevitable and the identified parameters are expected to depend on the input level.

A prior characterization step using multisine excitation with detection lines \cite{Schoukens2002} indicated that both odd and even nonlinearities are relevant. Therefore, monomials of degree 2 and 3 are chosen as basis functions. Increasing the monomial order was not considered since this would increase the number of coefficients and thus the computational effort significantly \cite{Paduart2010}.
The measured excitation force is determined as the input and the acceleration at the two sensor locations as the outputs of the identified state-space model. Two states are sufficient to describe the system's dynamics around the isolated resonance. Due to the nonlinearities in the output equation, \ie $\noncoeffy \neq \bm 0$, the model can capture changes in the deflection shape. 
With the degree of the polynomials, the number of outputs, inputs and states, there are 64 nonlinear coefficients.

To gather training data for the identification, the system was excited from 15 to 35 Hz with multisine signals at three different excitation levels. To this end, multisine signals with five realizations of the random phases were drawn for the lowest level and ten realizations for the other two levels.
Note that the constant amplitude spectrum holds for the input to the exciter. 
Due to the dynamics of the excitation system and the feedback of the structure to the excitation system, the amplitude spectrum of the excitation force is not constant (see \fref{MagBeam_excitation_spectra}).
The excitation forces have \RMS amplitudes of 0.5 N, 2.6 N and 5.2 N, respectively. The sampling frequency was 800 Hz, which is about 22.9 times higher than the highest excitation frequency.

\begin{figure} % PNLSS model mixed
	\centering
	\includegraphics{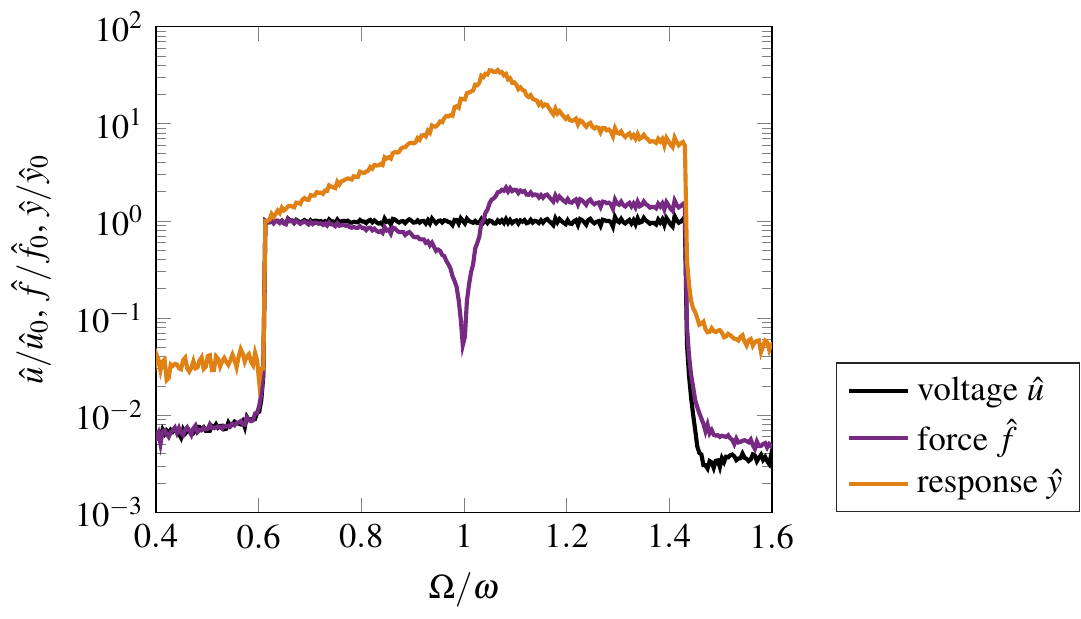}
	\caption{Normalized amplitude spectra of the voltage input to the shaker, the measured excitation force and the measured response of the structure.}
	\label{fig:MagBeam_excitation_spectra}
\end{figure}

The models are identified using four (lowest level) and nine (other two levels) realizations of measured training data of the same level. 
Based on different combinations of the realization of the training data, 20 models were identified for the lowest level and 30 models of the other two levels. Additionally, 30 other models are identified, called mixed models, using concatenated training data of all three levels.
No additional weighting is applied to account for different amplitude levels.

To compare the quality of the different \PNLSS models, the error between the simulated response $y_{\text{sim},i}$ and a measured (true) output $y_{\text{test},i}$ is computed. Note that only the $i$-th output is considered for this study. First, the true output is from a different realization of an input signal with the same amplitude spectrum and frequency band as the training data of the model. The input of that realization is used to obtain the simulated response $y_{\text{sim},i}$. This error is called same-level error. The true response and the absolute error of one of the models trained at mixed levels is shown both in time and frequency domain in \fref{MagBeam_models_pnlss}. Note that for this model, a realization of the highest amplitude spectrum was chosen as test data.
The error is about 5 \% of the vibration level in the frequency spectrum, except at about 24.6 Hz, just below the frequency with maximum response. Here, the error shows a peak with 14 \% of the vibration level.
Next, we compute the same-level error for all models. To better compare different models, the \RMS value of the error measure is normalized with the \RMS value of $y_{\text{test},i}$,
\ea{\pnlsserror = \dfrac{\sqrt{\frac{1}{N} \sum_{k = 1}^{N} (y_{\text{test},i}(\timevar_k) - y_{\text{sim},i}(\timevar_k) )^2}} {\sqrt{\frac{1}{N} \sum_{k = 1}^{N} y_{\text{test},i}(\timevar_k)^2}}.}{pnlss_error}
The mean of the relative error for each level is listed in the top row of \tref{MagBeam_pnlss_error}. 
The standard deviation of the error is stated in brackets. Note that the error is not normally distributed.
The models trained at 2.6 N \RMS have the lowest errors. The models trained at 5.2 N \RMS and mixed levels are labeled "unstable" since at least one of the models predicts unbounded amplitudes for the test data. Evidently, the identification of these models is highly dependent on to the training data. It seems that \PNLSS models identified for large excitation levels dependent more on the input. Large excitation levels lead to large amplitudes, where the influence of nonlinearity is more significant. This increases the significance of modeling errors. 

\begin{figure} % PNLSS model mixed
	\centering
	\begin{subfigure}{0.33\textwidth}
		\centering
		\includegraphics{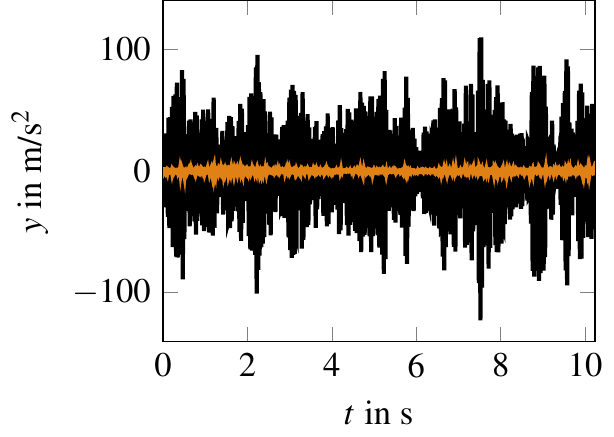}\caption{}	
	\end{subfigure}
	\begin{subfigure}{0.65\textwidth}
		\centering
		\includegraphics{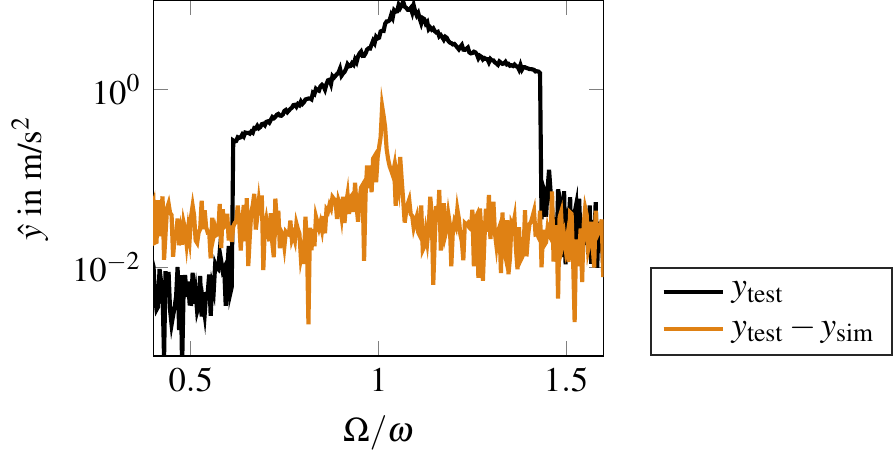}\caption{}	
	\end{subfigure}
	\caption{Absolute error $y_{\text{test},i} - y_{\text{sim},i}$ in (a) time and (b) frequency domain of the model trained at mixed levels compared with the test data set $y_{\text{test},i}$ of the same amplitude spectrum.}
	\label{fig:MagBeam_models_pnlss}
\end{figure}

It is important to note that the same-level error $\pnlsserror^\text{(same-level)}$ evaluates how well each model predicts the response to input data having the \textit{same} type and specifics as the training data. Training and input data are both multisines with the same amplitude spectrum. They merely have a different phase spectrum (randomly drawn).
For our study, it is useful to evaluate how well the models predict responses of a different amplitude spectrum than the spectrum they were trained on. To this end, the response to an input of the lowest (0.5 N \RMS) and the highest (5.2 N \RMS) amplitude spectrum is computed for all models. For both levels, all models are tested on five measured realizations for the low level and ten measured realizations of the high level.
The mean errors, $\pnlsserror^\text{(0.5 N \RMS)}$ for low-level input data and $\pnlsserror^\text{(5.2 N \RMS)}$ for high-level, as well as the range of the errors are also shown in \tref{MagBeam_pnlss_error}. All models predict bounded responses to data of lower level. However, if the models are excited with data of a higher amplitude spectrum than their training data, the polynomials are extrapolated, and the models generally do not perform well. This means that at least one model predicts unbounded responses for at least one realization of the test data. 

\begin{table} % table with validation errors
	\centering
	\begin{tabular}{ l m{2.5cm} m{3cm} m{3cm} m{3cm}}
		\toprule
		 & \multicolumn{4}{c}{Training data set}\\
		 & 0.5 N \RMS \newline (20 models) & 2.6 N \RMS \newline (30 models) & 5.2  N \RMS \newline (30 models)  & mixed \newline (30 models)\\
		\midrule[0.8pt] 
		\multicolumn{1}{l|}{$\pnlsserror^\text{(same-level)}$} &  6.3 \% (5.1 \%)  & 1.3 \% (0.3 \%) & unstable &  unstable \\
	    \multicolumn{1}{l|}{$\pnlsserror^\text{(0.5 N \RMS)}$} & (same-level)  & 5.6 \%  (4.8 \%) & 15.2 \%  (11.1 \%) &  6.5 \%  (4.7 \%) \\
		\multicolumn{1}{l|}{$\pnlsserror^\text{(5.2 N \RMS)}$} & unstable  & unstable & (same-level) &  unstable \\
		\bottomrule
		& \multicolumn{4}{c}{Selected models (one model for each training data set)}\\
		\midrule[0.8pt]  
		\multicolumn{1}{l|}{$\pnlsserror^\text{(same-level)}$} & 3.7 \%  & 1.2 \%  & 8.0 \% &  3.0 \% \\
		\multicolumn{1}{l|}{$\pnlsserror^\text{(10 N \RMS)}$} & (same-level) & 5.6 \% & 8.1 \% &  6.8 \%  \\
		\multicolumn{1}{l|}{$\pnlsserror^\text{(50 N \RMS)}$} & unstable & unstable & (same-level) & unstable \\
		\bottomrule
	\end{tabular}
	\caption{Mean of the relative errors of the different \PNLSS models according to \eref{pnlss_error}. The values in the brackets refer to the standard deviation of the error. The models are tested on data of the same amplitude spectrum as the training data ($\pnlsserror^\text{(same-level)}$), on data with the lowest amplitude spectrum $\pnlsserror^\text{(0.5 N \RMS)}$, and with the highest amplitude spectrum $\pnlsserror^\text{(5.2 N \RMS)}$. If at least one model predicts unbounded amplitudes for at least one realization of the test data, the cell is labeled "unstable". The upper part of the table is a statistical analysis of  several models, the lower part states the errors of selected models for each training data level.} \label{tab:MagBeam_pnlss_error}
\end{table}

\subsubsection{Nonlinear-Mode Model}

The backbone of the system was measured between 0.026 N and 0.56 N force amplitude, corresponding to the amplitude of the fundamental harmonic, with a sampling rate of 5000 Hz, \ie 205 times the linear modal frequency. The details on the controller settings are given in \aref{setup_parameters}.
The identified modal properties depending on the amplitude of the fundamental harmonic, $\hat{y}_1$, are shown in \fref{MB_freq_damp_mac}. The modal frequency increases by 3.7 \%, revealing the stiffening effect of the magnetic force.
The modal damping ratio increases from 0.4 \% to 0.45 \% with increasing amplitude and seems to saturate for high amplitudes. Interestingly, the identified linear modal damping ratio of the first mode in \tref{MagBeam_linearmodal} is equal to the average of the nonlinear damping shown in \fref{MB_freq_damp_mac} over this range of amplitude. The deviation between low-level nonlinear damping $\Dmod$ and the EMA damping $\zeta$ could be due to some inaccuracies in the EMA.
The amplitude dependence of the deflection shape is analyzed using the MAC value between the fundamental harmonic of the nonlinear deflection shape $\shpmodnorm_1$ and the linear mode shape $\shpmodlin$ identified with EMA,
\e{\text{MAC} = \frac{|\shpmodlin\tra \shpmodnorm_1 |^2}{\left( \shpmodlin\tra \shpmodlin \right) \left( \shpmodnorm\herm_1 \shpmodnorm_1 \right)}.}{MAC}
The MAC value decreases with amplitude but is close to one for all tested amplitudes, indicating that the fundamental harmonic of the deflection shape does not change significantly. 

\begin{figure} % modal model properties
	\centering
	\begin{subfigure}{0.49\textwidth}
		\centering
		\includegraphics[width=0.9\textwidth]{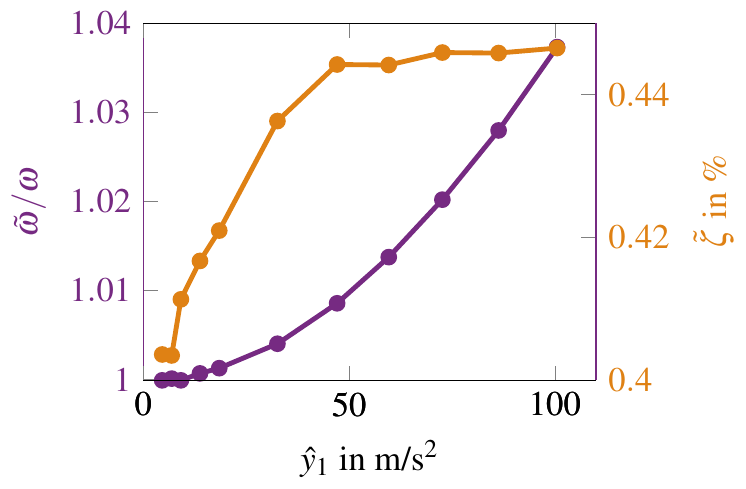}\caption{}
	\end{subfigure} 
	\begin{subfigure}{0.49\textwidth}
		\centering
		\includegraphics[width=0.75\textwidth]{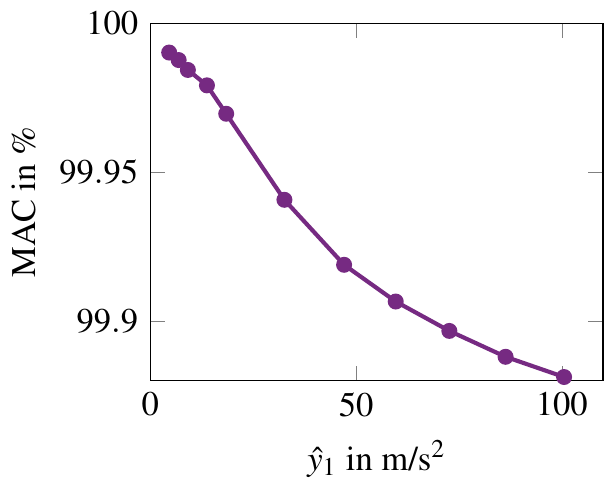}\caption{}	
	\end{subfigure}
	\renewcommand\figurename{Figure}
	\caption{(a) Identified modal frequency $\ommod$ and modal damping ratio $\Dmod$. (b) MAC value between the linear deflection shape and the fundamental harmonic of the nonlinear deflection shape. Note that the amplitude $\hat{y}_1$ corresponds to the fundamental harmonic and that the modal frequency is normalized with the first mode's linear modal frequency $\omega$ (c.f.\ \tref{MagBeam_linearmodal}).}
	\label{fig:MB_freq_damp_mac}
\end{figure}

The measurements are repeated four times, directly one after another, and good repeatability is observed with deviations in the modal frequency of different measurements of less than 0.1 \% for the same response level. 

\subsection{Prediction of Near-Resonant Vibrations}\label{sec:predict_MagBeam}

Next, the capability of the identified models for vibration prediction of two different excitation types is investigated: steady-state harmonic excitation and sine sweeps through resonance. 

\subsubsection{Steady-state Vibrations under Harmonic Excitation}

As reference, the system was excited with stepped sine signals. To this end, the phase lag between excitation force and drive-point acceleration was controlled using the PLL and varied around the resonance. Note that this is not the common stepped sine excitation where the excitation frequency is set. By controlling the phase lag, it is possible to stabilize and measure unstable branches of the frequency response \cite{Mojrzisch2012}. 
The \RMS level of the excitation force was controlled, and varied in a large range to obtain the different frequency responses. Only two representative levels are discussed in the following, namely 0.25 N and 0.55 N. 

For the prediction of steady-state vibrations, one representative \PNLSS model for both lowest and medium level are selected with the errors stated in the lower part of \tref{MagBeam_pnlss_error}. Additionally, one model of the highest level and one mixed model are selected, which predict bounded amplitudes for the same-level test data.
To fully capture the frequency response including unstable branches, harmonic balance is applied, including an alternating time-frequency scheme and path continuation to the time-discrete \PNLSS model, which is implemented in the open-source MATLAB tool NLvib \cite{Krack2019}. Only the fundamental harmonic of the excitation is considered for the vibration prediction, in the cases of both the \PNLSS model and the nonlinear-mode model \cite{Peter2018}. 
In the steady state ($\dot{\modamp} = 0 = \dot{\theta}$), the ordinary differential equation system \eref{slowflow} simplifies to the implicit equation
\ea{[- \Omega^2+2 \ii \Omega \ommod(\modamp) \Dmod(\modamp) +\ommod(\modamp)^2] \modamp \ee^{\ii\theta} = \shpmodnorm\herm_1 (\modamp) \ForceVec_{1,\rm exc},
}{nmsdof}
with excitation frequency $\Omega$. This equation is solved explicitly for $\Omega(\modamp)$ \cite{Schwarz2019}, with interpolated modal properties using piecewise cubic Hermite polynomials to achieve a finer resolution.

\begin{figure} % MagBeam steady-state synthesis
	\centering
	\begin{subfigure}{0.8\textwidth}
		\includegraphics[width=\textwidth]{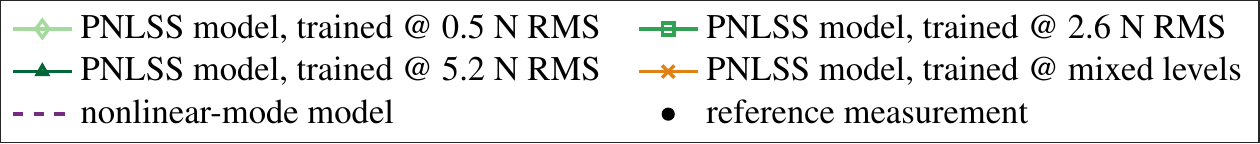}
	\end{subfigure} 
	\begin{subfigure}{0.45\textwidth}
		\includegraphics[width=\textwidth]{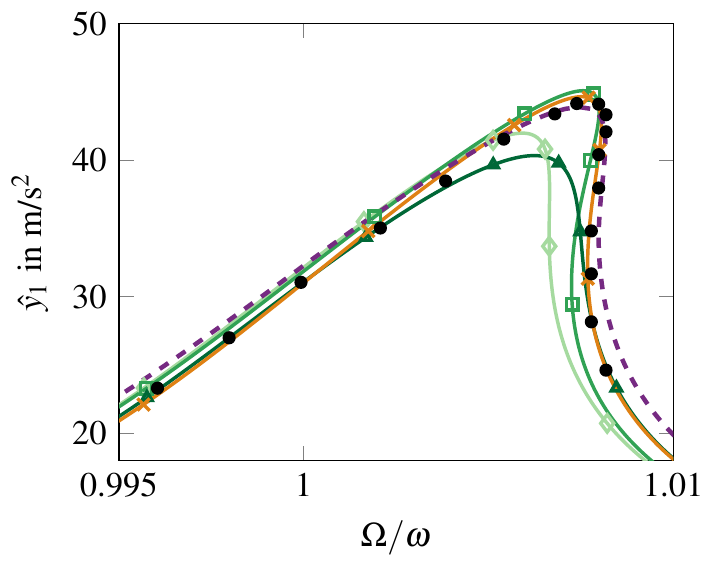}\caption{}
	\end{subfigure} 
	\begin{subfigure}{0.45\textwidth}
		\includegraphics[width=\textwidth]{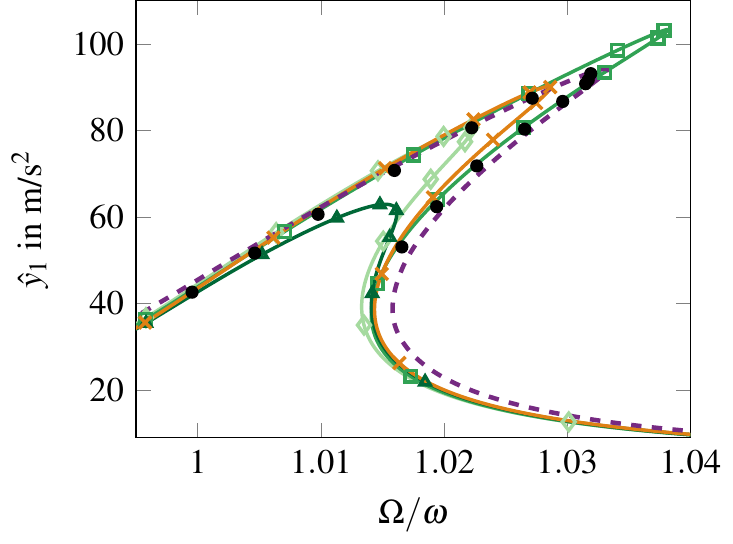}\caption{}
	\end{subfigure} 
	\renewcommand\figurename{Figure}
	\caption{Vibration prediction of steady-state frequency responses for (a) 0.25 N and (b) 0.55 N excitation level. The shown response amplitude is the magnitude of the fundamental harmonic.}	\label{fig:MB_steppedphase_amfreq}
\end{figure}

Both modal and \PNLSS models capture well the general shape of the frequency response, including turning points (see \fref{MB_steppedphase_amfreq}). However, the prediction accuracy of the different models regarding modal frequency and maximal amplitude differ significantly. 
The nonlinear-mode model predicts vibrations that are very close to the measured response for both excitation levels.
Of all \PNLSS models, the model trained at 2.6 N \RMS and at mixed levels yield the best predictions, which are equally good for the low excitation level. For the higher excitation level, the  model trained at 2.6 N \RMS overestimates the amplitudes, while the model trained at mixed levels underestimates them.
The models trained at 0.5 N \RMS and 5.2 N \RMS severely underestimate the amplitudes and the modal frequency in both cases (up to 30 \%). Intuitively, one would expect the \PNLSS model trained at 5.2 N \RMS to yield better predictions than a \PNLSS model trained at lower excitation levels, especially for high excitation amplitude. However, this is not the case here. Recall at this point that the training data of the \PNLSS models and the harmonic excitation is fundamentally different.

To further study the vibration prediction quality of \PNLSS models, two different mixed models are compared directly for the higher steady-state excitation level. One model is the mixed model of the selected models and the other mixed model is a model that predicts unbounded amplitudes for the same-level test data. The predicted steady-state frequency responses in \fref{MagBeam_steady_mixed} are similar. In fact, the "unstable" mixed model predicts the frequency response accurately. Thus, the same-level error is not a useful criterion to evaluate the steady-state predictions.

\begin{figure} 
	\centering
	\includegraphics[width=0.45\textwidth]{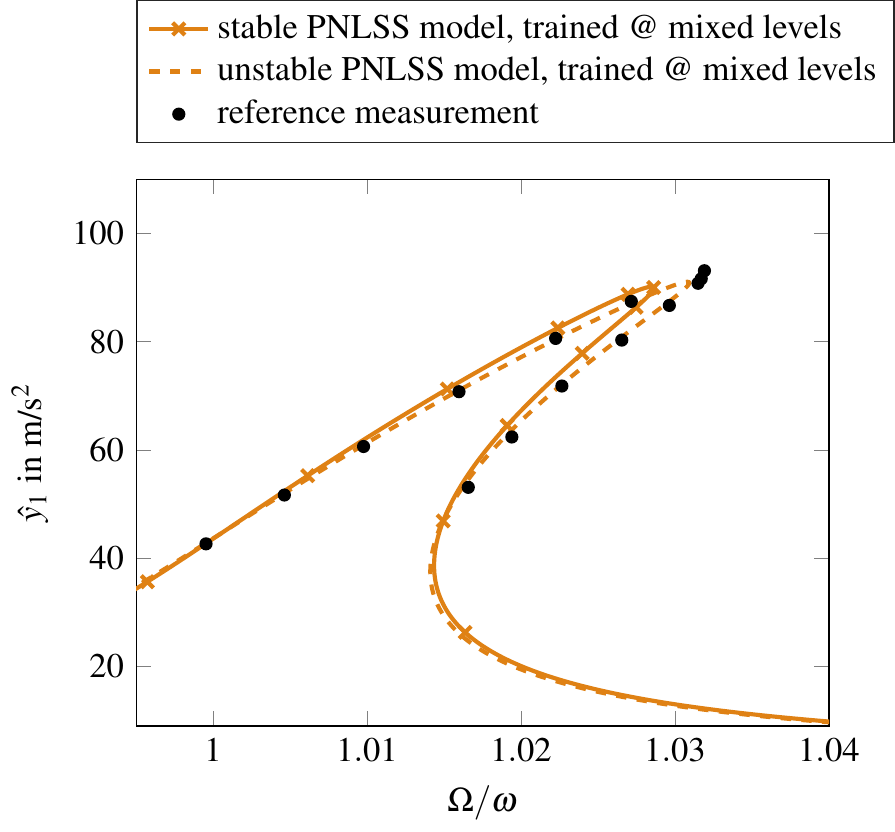}
	\renewcommand\figurename{Figure}
	\caption{Vibration prediction of steady-state frequency responses for 0.55 N excitation level. Comparison of two mixed models: one already shown in \fref{MB_steppedphase_amfreq} (marked with crosses), and one that predicts unstable vibrations when tested with same-level multisine test data.}	\label{fig:MagBeam_steady_mixed}
\end{figure}

\subsubsection{Transient Vibrations under Swept Sine Excitation}\label{sec:predict_MagBeam_transient}

As a second test case, sine sweeps through resonance are investigated. 
In this work, sweeps from 10 to 40 Hz are predicted for two different sweep rates: Slow sweeps with 0.25 Hz/s and fast sweeps with 7.5 Hz/s. The sweep rates are associated with a 1 \% frequency shift in the course of 23.75 and 0.79 pseudo-periods, respectively, where the pseudo-period is defined with the undamped linear modal frequency.
The sweeps are measured at three nominal shaker input levels for increasing frequencies only. Since the excitation force is not controlled, the actual force amplitude varies over time and shows a characteristic force drop at resonance (see \fref{MB_sweep_syn}a-c), consistent with the force drop seen in \fref{MagBeam_excitation_spectra}. Moreover, the shape of the envelope changes with different excitation level. Note that the variation in amplitude for the fast rate is no longer slow compared to the oscillation, which violates the main assumption of slow-flow averaging (inherent to the derivation of the nonlinear-mode model). The excitation force amplitude for low frequencies (before resonance) for the three levels are approximately 0.4 N, 4.3 N and 6.5 N. %The nominal amplitude to the shaker was 0.01 V, 0.1 V, and 0.15 V.
Note that all signals in \fref{MB_sweep_syn} are plotted for the number of pseudo-periods $N_{\text{periods}}$.

\begin{sidewaysfigure} % MagBeam transient force example
	\centering
	\begin{subfigure}{0.32\textwidth}
		\centering
		\includegraphics[width=\textwidth]{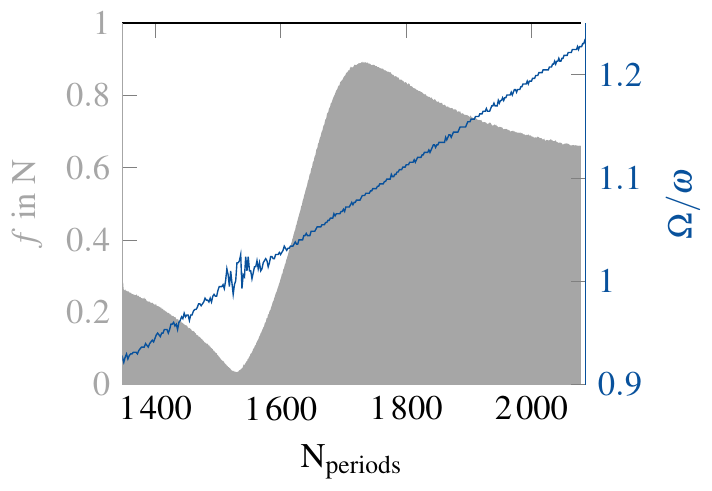}\caption{}
	\end{subfigure} 
	\begin{subfigure}{0.32\textwidth}
		\centering
		\includegraphics[width=\textwidth]{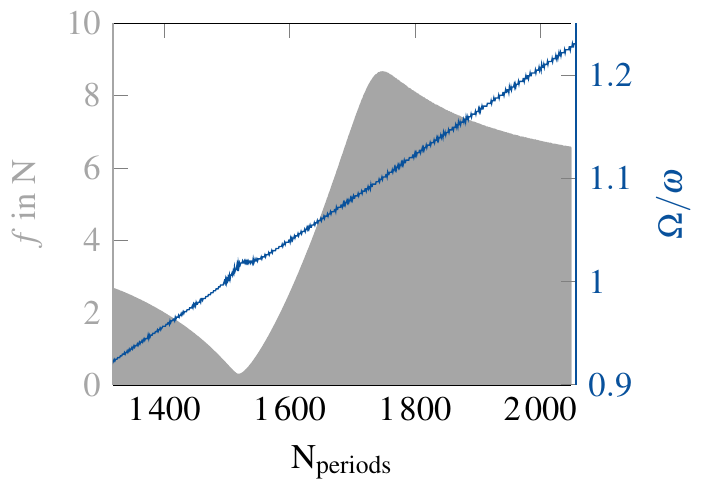}\caption{}
	\end{subfigure} 
	\begin{subfigure}{0.32\textwidth}
		\centering
		\includegraphics[width=\textwidth]{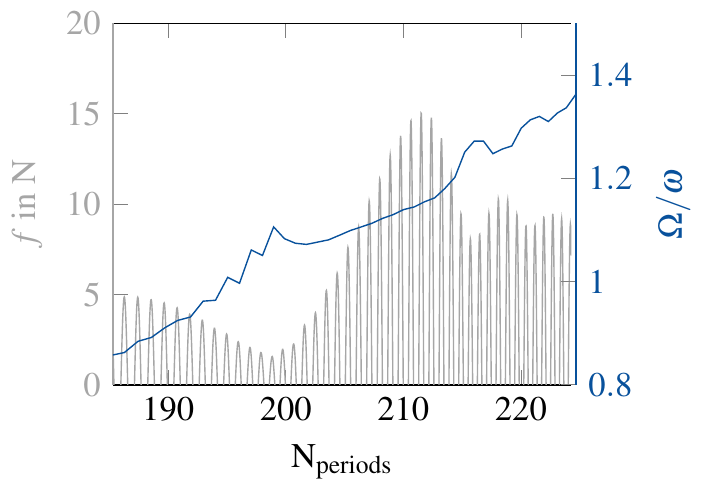}\caption{}
	\end{subfigure}
	\begin{subfigure}{0.32\textwidth}
		\centering
		\includegraphics[width=0.8\textwidth]{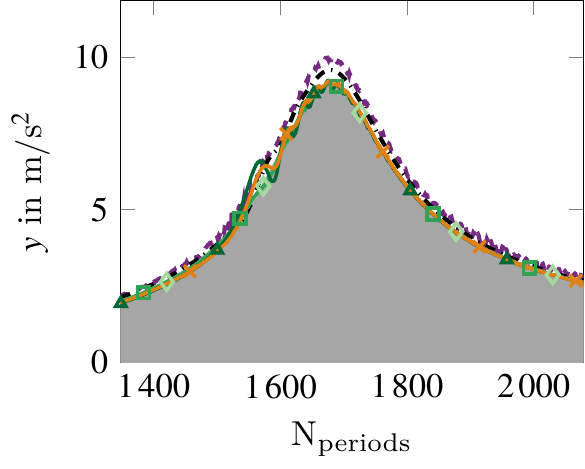}\caption{}
	\end{subfigure} 
	\begin{subfigure}{0.32\textwidth}
		\includegraphics[width=0.8\textwidth]{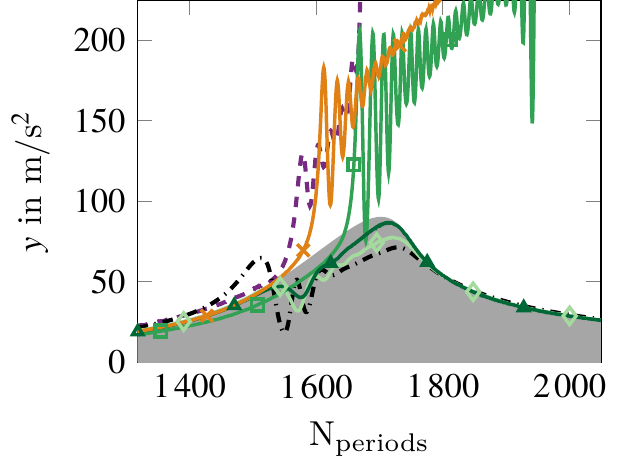}\caption{}		
	\end{subfigure}
	\begin{subfigure}{0.32\textwidth}
		\includegraphics[width=0.8\textwidth]{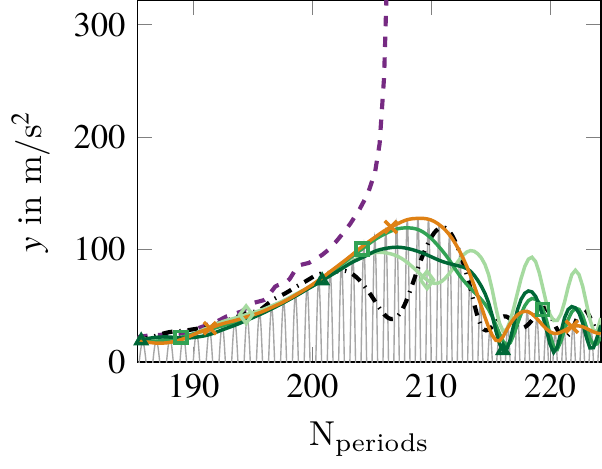}\caption{}
	\end{subfigure} 
	\begin{subfigure}{0.8\textwidth}
		\includegraphics[width=\textwidth]{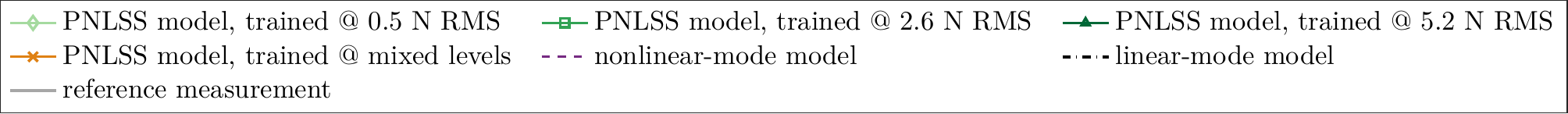}
	\end{subfigure} 
	\caption{(a)-(c) Measured excitation force and instantaneous excitation frequency. (d)-(f) Predicted response. The excitation level of sweep (a)/(d) is 0.4 N, of sweep (b)/(e) is 4.3 N and of sweep (c)/(f) is 6.5 N. The sweep rate of sweeps (a)/(d) and (b)/(e) is 0.25 Hz/s and of sweep (c)/(f) is 7.5 Hz/s.}	\label{fig:MB_sweep_syn}
\end{sidewaysfigure}

To predict the vibration with \PNLSS models, the measured force signal is used as an input signal to the four representative \PNLSS models. For the nonlinear-mode model, the slow-flow averaging algorithm requires time continuous functions for the instantaneous frequency and amplitude of the force signal (\cf \eref{slowflow}). To this end, the instantaneous frequency, shown in \fref{MB_sweep_syn}a-c, is detected via the time difference between zero crossings. 
Note that even though the sweep rate is constant for the input voltage signal to the shaker, the instantaneous frequency extracted from the force signal does not show a linear increase, which is due to shaker-structure interactions.
For a cross-check, the extraction of instantaneous magnitude and frequency was determined using a continuous wavelet transform with Morlet wavelets. No significant differences in the two analyses were found. Therefore, we preferred the approach of zero-crossing detection due to its simplicity. 
The maximum force value between two zero crossings is defined as the instantaneous magnitude. Note that this is an approximation of the required fundamental harmonic $\ForceVec_{1,\rm exc}$ in \eref{slowflow}. The extracted data is interpolated with piecewise cubic Hermite polynomials to obtain time-continuous functions. 

For the lowest excitation level, the \PNLSS models trained at 0.5 N \RMS and 2.6 N \RMS predict the vibration well, whereas model trained at 5.2 N \RMS and mixed levels show small envelope modulation for frequencies lower than the modal frequency. 
To asses the gain of using \PNLSS models compared to a linear model, a single-degree-of-freedom modal model is set up based on the first mode's linear modal properties, identified with EMA (see \tref{MagBeam_linearmodal}). Recall that this linear model should be valid for low-level responses and its modal frequency is lower than in the nonlinear case. The linear model overestimates the amplitudes. The nonlinear-mode model predicts even higher amplitudes, since the nonlinear damping is lower than linear damping for low energies (\cf \fref{MB_freq_damp_mac} and \tref{MagBeam_linearmodal}).
For the second excitation level, no model predicts the vibration with acceptable accuracy. The responses of the nonlinear-mode model and \PNLSS model trained at 2.6 N \RMS are unstable, \ie unbounded amplitudes are predicted. The predicted amplitudes of the mixed \PNLSS model stay bounded but are too high by a factor of 6. The \PNLSS models trained at 0.5 N \RMS, 5.2 N \RMS, and the linear model underestimate the resonance amplitudes and show significant modulations in the envelope prior to the resonance.
For the highest excitation level with the fast sweep rate, the \PNLSS model trained at mixed levels accurately predicts the response. The other \PNLSS models as well as the linear model underestimate the maximal amplitude and show modulations. The nonlinear-mode model predicts unbounded amplitudes.

As previously, two mixed models are compared directly for the transient sine sweeps (b)/(e) and (c)/(f) (see \fref{MagBeam_transient_mixed}). Both models predict very similar responses, meaning both predict too high amplitudes for sweep (b)/(e) and predict accurately the sweep (c)/(f).

\begin{figure} % setup
	\centering
	\begin{subfigure}{0.8\textwidth}
		\centering
		\includegraphics[width=0.6\textwidth]{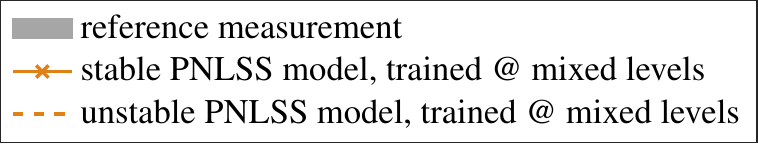}
	\end{subfigure} 
	\begin{subfigure}{0.49\textwidth}
		\centering
	\includegraphics[width=0.8\textwidth]{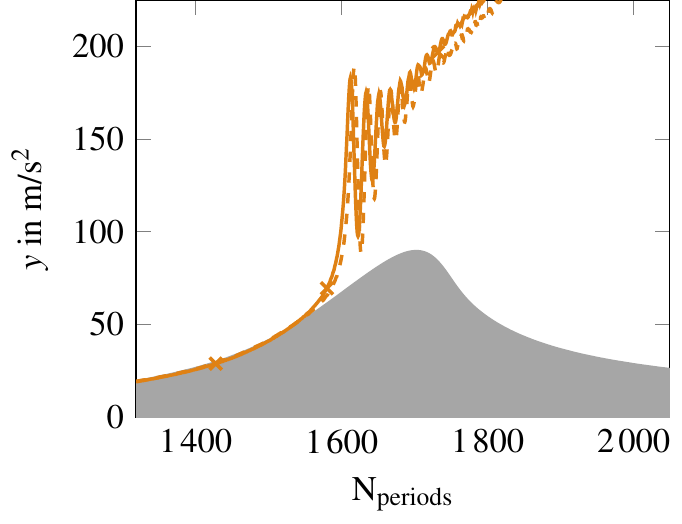}
		\caption{}
	\end{subfigure} 
	\begin{subfigure}{0.49\textwidth}
		\centering
	\includegraphics[width=0.8\textwidth]{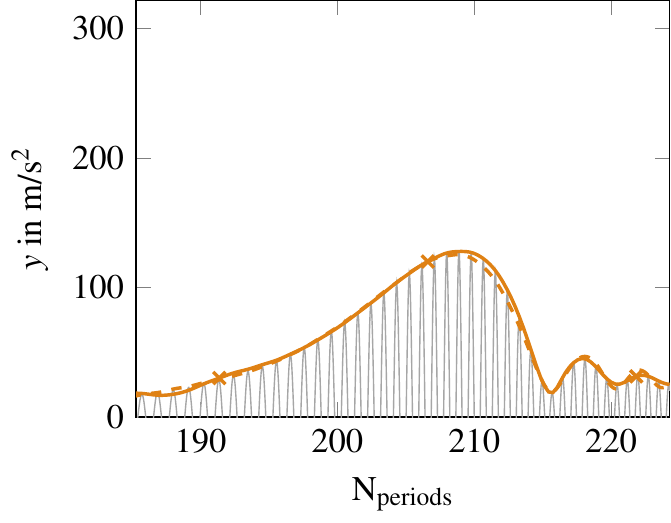}
		\caption{}
	\end{subfigure} 
	\renewcommand\figurename{Figure}
	\caption{Sine sweep prediction of the sweeps \fref{MB_sweep_syn} (b)/(e) and (c)/(f) for two different mixed models: one already shown in \fref{MB_sweep_syn} (marked with crosses), and one that predicts unstable vibrations when tested with same-level multisine test data.}
	\label{fig:MagBeam_transient_mixed}
\end{figure}

To explain the instability predicted by the nonlinear-mode model, a numerical study was performed. One key feature of the measurements is the force drop and subsequent overshoot of the excitation force between 1500 and 1700 or 200 and 210 periods for the slow and fast sweeps, respectively. Due to the shaker-structure interaction, the vibration response influences the excitation force. In the nonlinear modal synthesis, this shaker-structure interaction is ignored, instead only the measured force is imposed. 
Consequently, small errors in modal frequency result in much higher excitation level at the (incorrect) modal frequency and therefore lead to overpredicted responses at those frequencies.
Higher response amplitudes lead to a higher modal frequency due to the stiffening nonlinearity in the system. Therefore, the resonance condition is maintained, resulting in a resonance capture, which is generally known for stiffening systems and upward sweeps. Additionally, at high response levels, extrapolation of the modal properties can lead to incorrect results.
With the described mechanism, a small model error (in terms of modal frequency) can lead to large simulation errors, which was replicated in the numerical study.

%%%%%%%%%%%%%%%%%%%%%%%%%%%%%%%%%%%%%%%%%%%%%%%%%%%%%%%%%%%%%%%%%%%%%%%%%%%%%%%%%%%%%%%%%%%%

\section{Application to a System of Two Beams with Bolted Joints (BRB)} \label{sec:BRB}

\subsection{Experimental Setup and Testing Procedure}

The second specimen is the \BRB, which is assembled from two beams, jointed by three bolts with washers \cite{Brake2019}. Note that during preparation of this article, selected results obtained with the BRB have been presented at a conference \cite{Scheel2018b}. The nonlinear contact interactions depend on the bolt torque \cite{Brake2018,Balaji2019}. In this work, a relatively small bolt torque of 5 Nm was chosen.
Note that the nonlinear interactions for this specimen, \ie local stick and slip and contact-separation events between rough surfaces \cite{Chen2019}, lead to a non-smooth nonlinear force-deformation relation, which cannot be well-approximated with low-order polynomials.
In this work, the beam was kept assembled between the measurements to exclude the well-known variability due to reassembly.

The beam was suspended with nylon strings and bungee cords approximately 30 mm away from each end of the beam. The excitation was applied using a shaker-stinger apparatus in the lateral direction with a 2 mm diameter steel rod of 65 mm length and the Br\"uel \& Kj\ae r Vibration Exciter 4808.
Four PCB 352A24 uniaxial accelerometers were glued to the beam and one PCB 288D01 impedance head was screwed to the beam at the locations indicated in \fref{BRB_setup}. In the following, all results will be shown for the sensor on the right side of the joint. 

\begin{figure} % setup
	\centering
	\begin{subfigure}{0.5\textwidth}
		\centering
		\def\svgwidth{1\textwidth}
		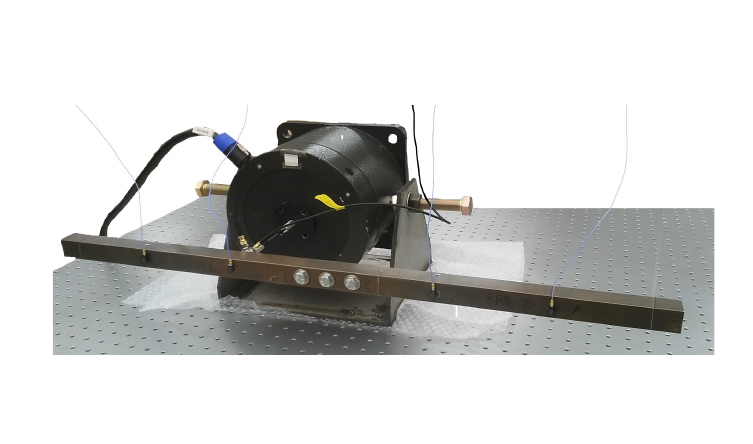\caption{}
	\end{subfigure} 
	\begin{subfigure}{0.8\textwidth}
		\centering
		\def\svgwidth{0.9\textwidth}
		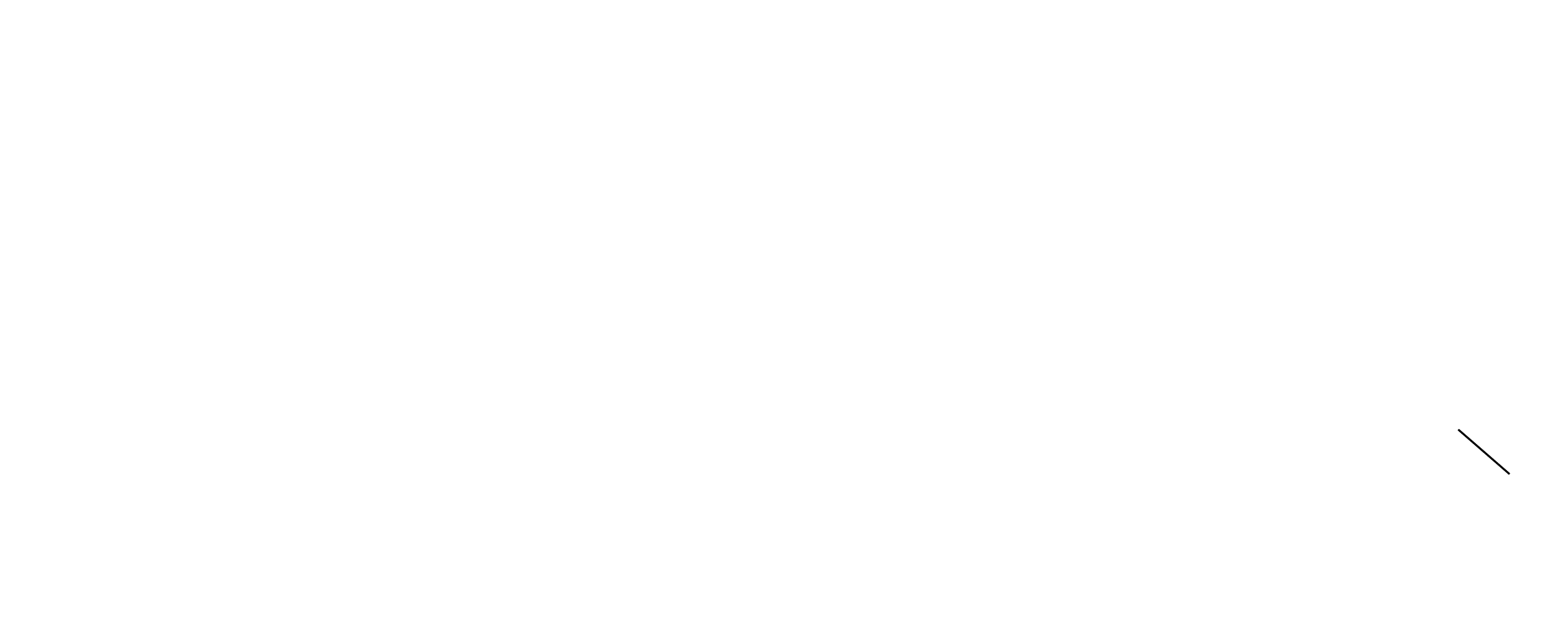\caption{}
	\end{subfigure} 
	\renewcommand\figurename{Figure}
	\caption{Picture (a) and scheme (b) of the BRB and the experimental setup.}
	\label{fig:BRB_setup}
\end{figure}

The measurement campaign for this specimen was the same as for the beam with magnets (see \tref{MagBeam_measurement}). 
A linear EMA was performed, exciting the frequency range from 100 to 2000 Hz with a pseudo-random multisine signal. Four well-separated modes were identified in this range (see \tref{BRB_linearmodal}). Only the first bending mode (mode 1) is investigated in this study.
\begin{table} % table with linear modal properties
	\centering
	\begin{tabular}{ l l l l l }
		\toprule
		& mode 1 & mode 2 & mode 3 & mode 4\\
		\midrule[0.8pt]  
		linear modal frequency $\omega$ & 149.0 Hz & 564.3 Hz & 1138.3 Hz & 1496.5 Hz\\
		linear modal damping ratio $\zeta$ & 0.14 \% & 0.05 \% & 0.08 \% & 0.11 \%\\
		\bottomrule
	\end{tabular}
	\caption{Linear modal frequency and damping ratio for the first four bending modes, identified with EMA at low level.}\label{tab:BRB_linearmodal}
\end{table}

\subsection{Model Identification}

\subsubsection{\PNLSS Models}

A prior characterization step indicated that both odd and even nonlinearities are relevant. Therefore, monomials of degree 2 and 3 are chosen as basis functions. The monomial order was not further increased for the reasons explained earlier (see \sref{BRB_PNLSS_models}). 
The identified \PNLSS models have one input and four outputs corresponding to the excitation force and the measured accelerations. Note that the acceleration recorded by the impedance head at the drive point was considered, discarding the data of the accelerometer on the opposite side.
No\"{e}l \etal \cite{Noel2017a} showed that three states are necessary to replicate the hysteresis dynamics of a single-degree-of-freedom oscillator. Due to the hysteretic nature of the nonlinearity for the \BRB and the antiresonance in the frequency range of interest, four states were chosen. With this, the number of nonlinear coefficients is 400.

Multisine excitation was applied to the beam around the first resonance from 75 to 225 Hz. The sampling frequency was 6400 Hz, which is about 28.4 times higher than the highest excitation frequency. Four excitation levels were tested, having approximately \RMS amplitudes of 10 N, 20 N, 30 N, and 50 N. The voltage amplitude spectrum is constant, but the measured excitation force varies in amplitude (see \fref{BRB_excitation_spectra}). Of each level, ten realizations were measured.

For each excitation level, 30 \PNLSS models were identified as well as 30 mixed models with concatenating data without additional weighting. 
In the following, only models trained at 10 N \RMS, 30 N \RMS, 50 N \RMS and mixed levels are shown as representative results.

\begin{figure} % PNLSS model mixed
	\centering
	\includegraphics{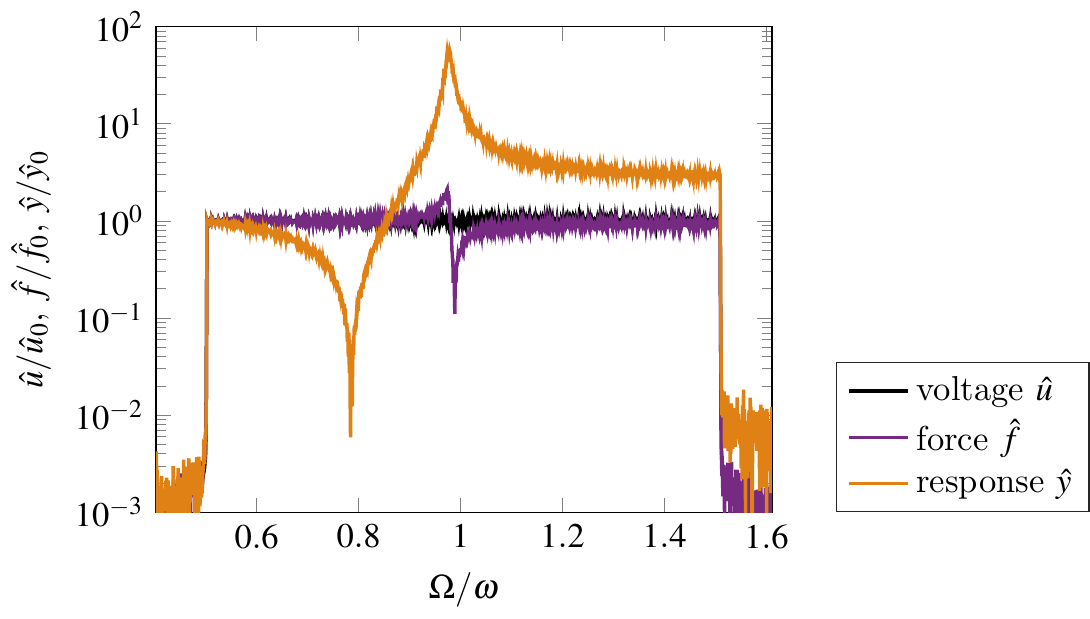}
	\caption{Normalized amplitude spectra of the voltage input to the shaker, the measured excitation force and the measured response of the structure.}
	\label{fig:BRB_excitation_spectra}
\end{figure}

As an example, one mixed model is tested on a realization of the highest excitation level (see \fref{BRB_models_pnlss}). The error is about 6 \% of the vibration level for the frequency with maximal response. At the antiresonance, the error is of similar level as the response.
To compare the identified models, first, the relative error measure $\pnlsserror^\text{(same-level)}$ for test data of the same amplitude spectrum as in the training data is computed for all models. The mean of the relative error for each level is stated in \tref{BRB_pnlls_error} with the standard deviation stated in brackets. Note that the error is not normally distributed.
For this specimen, some \PNLSS models trained at 50 N \RMS and mixed levels predict unbounded amplitudes. As previously, models trained at large amplitudes are more input-dependent, potentially due to the increased influence of modeling errors. It is clear that non-smooth nonlinearities caused by contact interactions at the interface are not correctly modeled by polynomials. By choosing only low-order polynomials, model form error is introduced into the model. 
The models are also tested with ten realizations of test data of lower level and higher level, respectively. For low-level test data, the prediction of all models stay bounded. For test data of the highest level, models trained at all possible levels can predict unbounded amplitudes.

\begin{figure} % PNLSS model mixed
	\centering
	\begin{subfigure}{0.33\textwidth}
		\centering
		\includegraphics{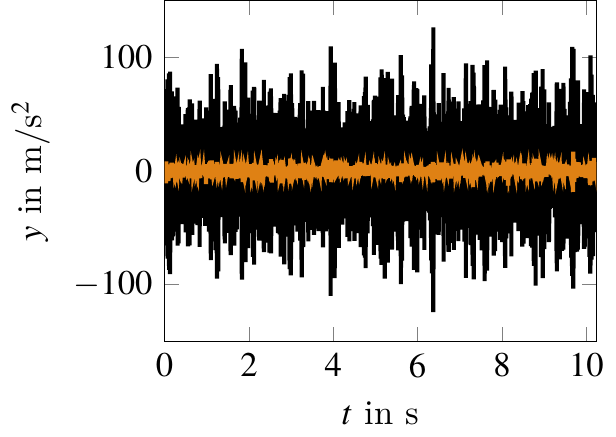}\caption{}
	\end{subfigure}
	\begin{subfigure}{0.65\textwidth}
		\centering
		\includegraphics{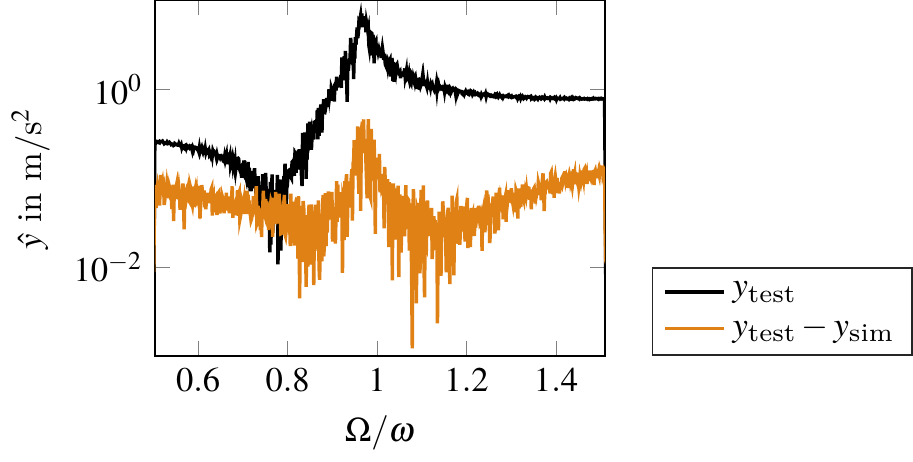}\caption{}
	\end{subfigure}
	\caption{Absolute error $y_{\text{test},i} - y_{\text{sim},i}$ in (a) time and (b) frequency domain of the model trained at mixed levels compared with a test data set $y_{\text{test}}$ fo the same amplitude spectrum.}
	\label{fig:BRB_models_pnlss}
\end{figure}

\begin{table} % table with validation errors
	\centering
	\begin{tabular}{ l m{2.5cm} m{3cm} m{3cm} m{3cm}}
		\toprule
		& \multicolumn{4}{c}{Training data set}\\
		 & 10 N \RMS \newline (30 models) & 30 N \RMS \newline (30 models) & 50  N \RMS \newline (30 models) & mixed \newline (30 models) \\
		\midrule[0.8pt]  
		\multicolumn{1}{l|}{$\pnlsserror^\text{(same-level)}$} & 8.9 \% (9.4 \%) & 5.1 \%  (1.6 \%) & unstable &  unstable \\
		\multicolumn{1}{l|}{$\pnlsserror^\text{(10 N \RMS)}$} & (same-level) & 7.2 \% ( 1.9 \%) & 30.1 \%  (5.0 \%) &  5.5 \%  (2.0 \%) \\
		\multicolumn{1}{l|}{$\pnlsserror^\text{(50 N \RMS)}$} & unstable & unstable & (same-level) & unstable \\
		\bottomrule
		& \multicolumn{4}{c}{Selected models (one model for each training data set)}\\
		\midrule[0.8pt]  
		\multicolumn{1}{l|}{$\pnlsserror^\text{(same-level)}$} & 2.7 \%  & 4.8 \%  & 7.3 \% &  8.2\% \\
		\multicolumn{1}{l|}{$\pnlsserror^\text{(10 N \RMS)}$} & (same-level) & 7.4 \%  & 31.5 \% &  8.6 \%  \\
		\multicolumn{1}{l|}{$\pnlsserror^\text{(50 N \RMS)}$} & unstable & unstable & (same-level) & unstable \\
		\bottomrule		
	\end{tabular}
	\caption{Mean of the relative errors of the different \PNLSS models according to \eref{pnlss_error}. The values in the brackets refer to the standard deviation of the error. The models are tested on data of the same amplitude spectrum as the training data ($\pnlsserror^\text{(same-level)}$), on data with the lowest amplitude spectrum $\pnlsserror^\text{(10 N \RMS)}$, and with the highest amplitude spectrum $\pnlsserror^\text{(50 N \RMS)}$. If at least one model predicts unbounded amplitudes for at least one realization of the test data, the cell is labeled "unstable". The upper part of the table is a statistical analysis of  several models, the lower part states the errors of selected models for each training data level.} \label{tab:BRB_pnlls_error}
\end{table}

\subsubsection{Nonlinear-Mode Model}

The modal model is identified for a backbone measured at eleven excitation levels from 0.1 N to 16.6 N amplitude of the fundamental harmonic with a sampling rate of 5000 Hz, \ie 33.6 times the linear modal frequency.
The modal frequency decreases by 2.2 \% over the amplitude range while the damping ratio increases from 0.17 \% to 1.35 \% (see \fref{freq_damp_mac}).
The MAC value (\cf \eref{MAC}) is close to one in the tested amplitude regime, which indicates that the deflection shape does not change significantly with increasing amplitude.

To assess repeatability, five backbones were measured, directly one after another. This was done in the same manner for all backbones, \ie with the same increasing nominal shaker input level. For points of the same shaker input level, the modal frequency was well repeatable (below 0.1 \% variation), but the force amplitude and the modal response amplitude exhibit variations of up to 7.5 \% and 1.3 \%, respectively, which then affects the extracted modal damping ratio.

\begin{figure} % modal properties
	\centering
	\begin{subfigure}{0.49\textwidth}
		\centering
		\includegraphics[width=0.9\textwidth]{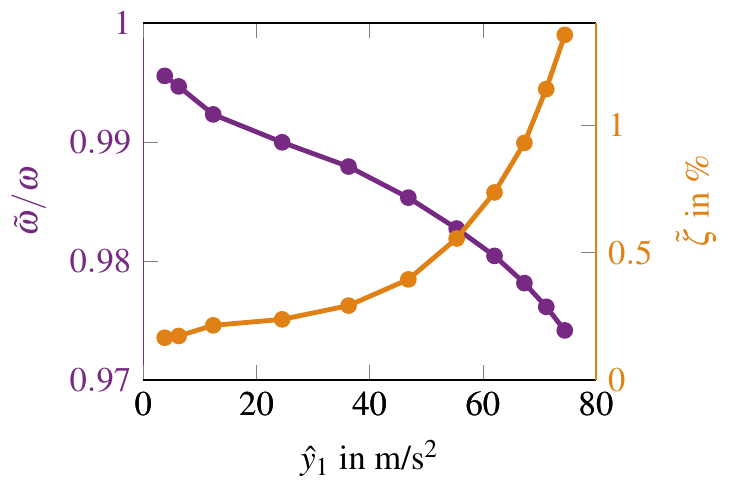}\caption{}	
	\end{subfigure} 
	\begin{subfigure}{0.49\textwidth}
		\includegraphics[width=0.85\textwidth]{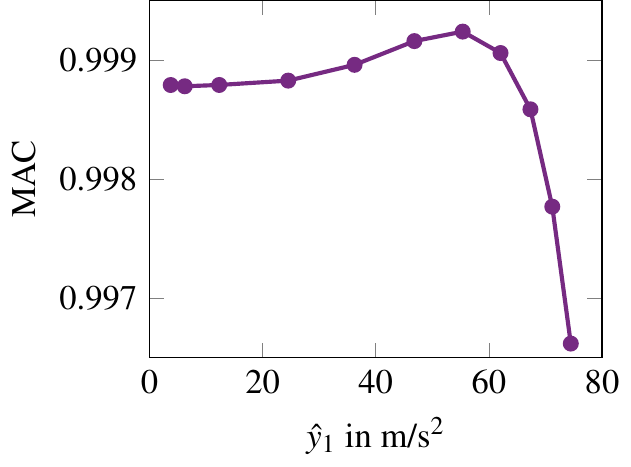}\caption{}	
	\end{subfigure}
	\renewcommand\figurename{Figure}
	\caption{(a) Identified modal frequency and modal damping ratio depending on the vibration amplitude. (b) MAC value between the linear deflection shape and the fundamental harmonic of the nonlinear deflection shape. Note that the amplitude corresponds to the fundamental harmonic.}
	\label{fig:freq_damp_mac}
\end{figure}

\subsection{Prediction of Near-Resonant Vibrations}\label{sec:predict_BRB}

\subsubsection{Steady-state Vibrations under Harmonic Excitation}\label{sec:predict_BRB_steady}

As reference measurements, sine measurements with amplitude and phase control of the excitation force were performed. Five levels were tested, starting at the highest excitation level. Each level was measured four times.
The averaged results are shown for three representative levels, namely for 2 N, 5 N, and 15 N (see \fref{BRB_steppedphase_amfreq}). The reference measurements were well repeatable: the frequency varied by less than 0.2 \% and the amplitudes varied by up to 7.4 \%.
The nonlinear-mode model predicts amplitudes smaller than the resonance for the excitation levels 2 N and 5 N. For the highest excitation level, the amplitudes and the modal frequency differ only slightly from the reference measurements. In theory, the phase-resonant point of the frequency response must lie on the backbone. However, interpolation errors or repetition-variability can lead to deviations in practice. Consequently, if one compares measured points of the backbone to corresponding points of the reference sine measurements, the response amplitudes in the reference measurements are larger than in the backbone measurement. All prediction errors must be discussed in the light of this variability which is typical for systems with joints, especially the \BRB, known for its inherent variability due to changes at the contact interface \cite{Brake2019}.

One representative \PNLSS model for each level was selected, with the errors stated in \tref{BRB_pnlls_error}.
These \PNLSS models significantly underestimate the maximum amplitude for the 2 N excitation level, where the predicted modal frequency differs between models. For the 5 N excitation level, all \PNLSS models but one significantly underestimate the maximum amplitude (in comparison to the repetition-variability). Only the \PNLSS model trained at 10 N \RMS predicts the amplitude for the 5 N excitation with less than 0.5 \% error. For the 15 N excitation, in contrast, the predicted response amplitudes of \PNLSS model trained at 10 N \RMS are too high by a factor of 4.
In addition to the inherent repetition-variability, the high variability of predictions of \PNLSS models stems from the fact that the identified model depends on the training data. The prediction quality does not clearly correlate with the testing error $\pnlsserror$ for this specimen and the selected models. The models trained at 30 N \RMS and 50 N \RMS, for example perform equally in the prediction but differ substantially in the error measure for all tested levels.  Moreover, no \PNLSS model provides a reasonable accuracy in the full range of excitation levels.

\begin{figure} % steady-state synthesis BRB
	\centering
	\begin{subfigure}{0.8\textwidth}
		\centering
		\includegraphics[width=\textwidth]{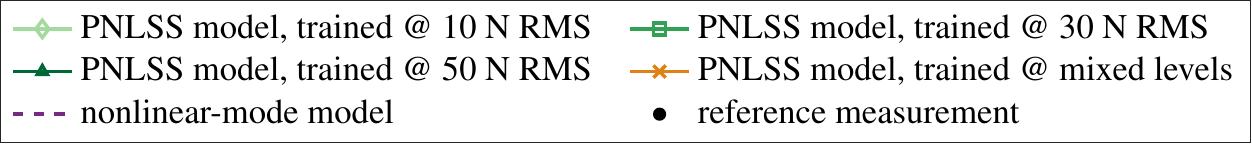}
	\end{subfigure} 
	\begin{subfigure}{0.32\textwidth}
		\centering
		\includegraphics[width=\textwidth]{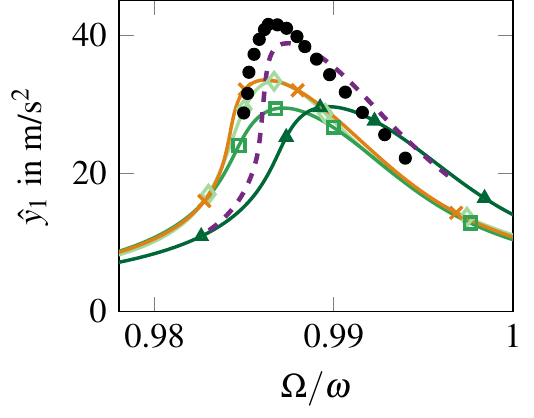}\caption{}
	\end{subfigure} 
	\begin{subfigure}{0.32\textwidth}
		\centering
		\includegraphics[width=\textwidth]{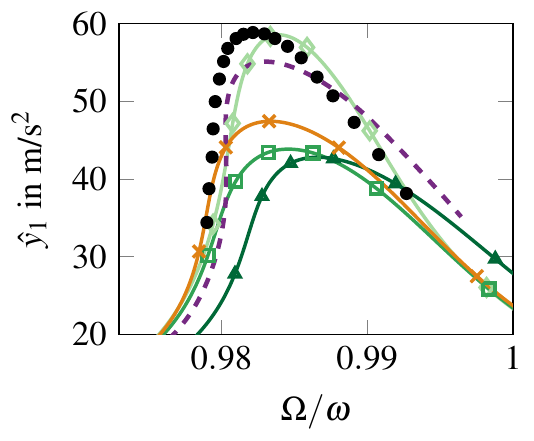}\caption{}
	\end{subfigure} 
	\begin{subfigure}{0.32\textwidth}
		\centering
		\includegraphics[width=\textwidth]{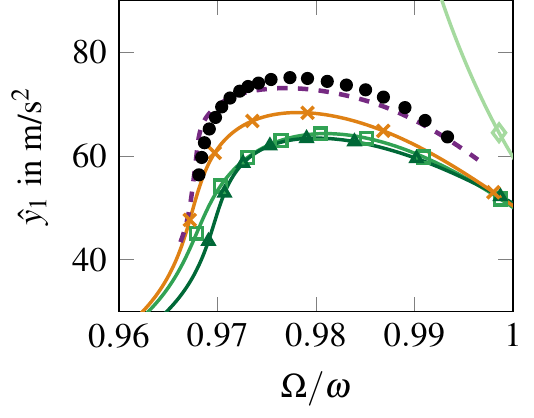}\caption{}
	\end{subfigure} 
	\renewcommand\figurename{Figure}
	\caption{Vibration prediction of steady-state frequency responses for (a) 2 N, (b) 5 N and (c) 15 N excitation level. The shown response amplitude is the magnitude of the fundamental harmonic.}	\label{fig:BRB_steppedphase_amfreq}
\end{figure}

To further assess the prediction variation of \PNLSS models, four mixed models are compared for the highest steady-state excitation level (see \fref{BRB_steady_mixed}). One of the models is the selected mixed model already shown in \fref{BRB_steppedphase_amfreq}. Additionally, one model that predicts bounded vibrations when tested with same-level multisine data (with $\pnlsserror^{(\text{same-level})} = 5.1 \%$) and two models that predict unstable vibrations when tested with same-level multisine test data are selected. The shape of the vibration predictions differ qualitatively, meaning that some models falsely predict right-bending frequency responses. These false predictions are independent of the value of $\pnlsserror^{(\text{same-level})}$.
As mentioned earlier, low-order polynomials apparently cannot replicate the physics of a joint, thus leading to these qualitatively wrong predictions.

\begin{figure} 
		\centering
		\includegraphics[width=0.5\textwidth]{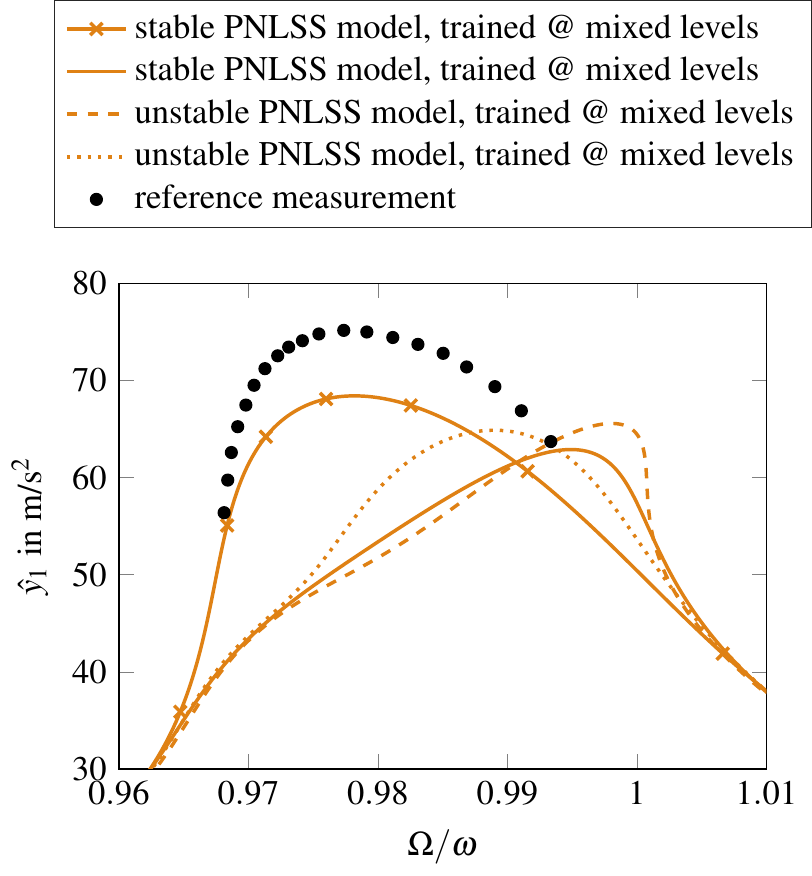}
	\renewcommand\figurename{Figure}
	\caption{Vibration prediction of steady-state frequency responses for 15 N excitation level. Comparison of four mixed models: one already shown in \fref{BRB_steppedphase_amfreq} (marked with crosses), another model that predicts bounded vibrations when tested with same-level multisine data and two models that predict unstable vibrations when tested with same-level multisine test data.}	\label{fig:BRB_steady_mixed}
\end{figure}

\subsubsection{Transient Vibrations under Swept Sine Excitation}\label{sec:predict_BRB_transient}

For the BRB, sine sweeps at six different excitation levels were measured with a sweep rate of 15 Hz/s from 130 to 190 Hz, corresponding to a 1 \% frequency shift in 14.81 pseudo-periods. The sweeps are uncontrolled, which lead to variation of the excitation force, especially to oscillations after resonance. All sweeps were measured four times with negligible variation between the measurements. The sweeps are shown in \fref{BRB_sweep_syn} for two representative nominal force levels with 8.1 N and 19.1 N force amplitude attained off resonance. Note that for the lower level, the instantaneous frequency drops around resonance, where the force is very low such that the noise has more influence and the zero crossing detection is not robust. The sampling rate was 12800 Hz. To obtain time-continuous instantaneous magnitude and frequency of the excitation force, the procedure described in \sref{predict_MagBeam_transient} is followed.

The accuracy of the nonlinear-mode model is moderate, \ie the predicted amplitudes are consistently higher than the reference. For the \BRB, the prediction does not suffer from the instability seen in the magnetic beam specimen.
The \PNLSS model trained at mixed levels is the best of all selected \PNLSS models. It accurately predicts the low-level sweep and has only up to 2.2 \% error of the maximal amplitude in the high-level sweep. \PNLSS model trained at 30 N \RMS predicts with similar accuracy. The amplitudes predicted with \PNLSS model trained at 10 N \RMS are higher than the reference. \PNLSS model trained at 50 N \RMS predicts lower amplitudes and more pronounced amplitude modulations after the resonance.

\begin{figure} % MagBeam transient force example
	\centering
	\begin{subfigure}{0.45\textwidth}
		\centering
		\includegraphics[width=0.95\textwidth]{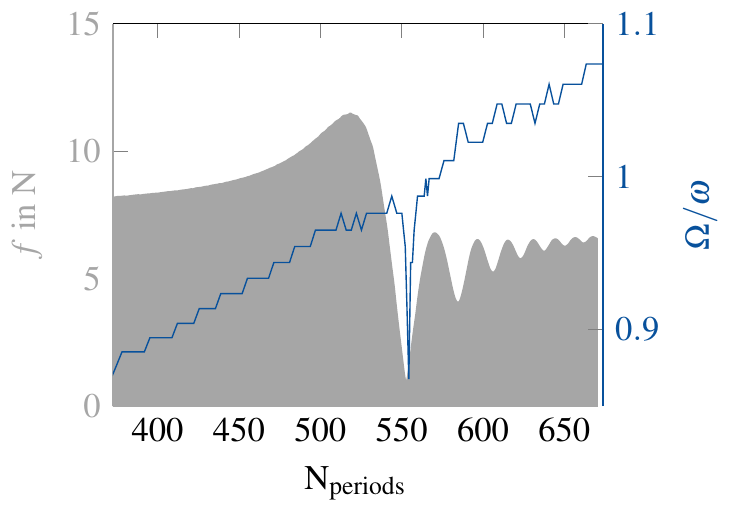}\caption{}
	\end{subfigure} 
	\begin{subfigure}{0.45\textwidth}
		\centering
		\includegraphics[width=0.9\textwidth]{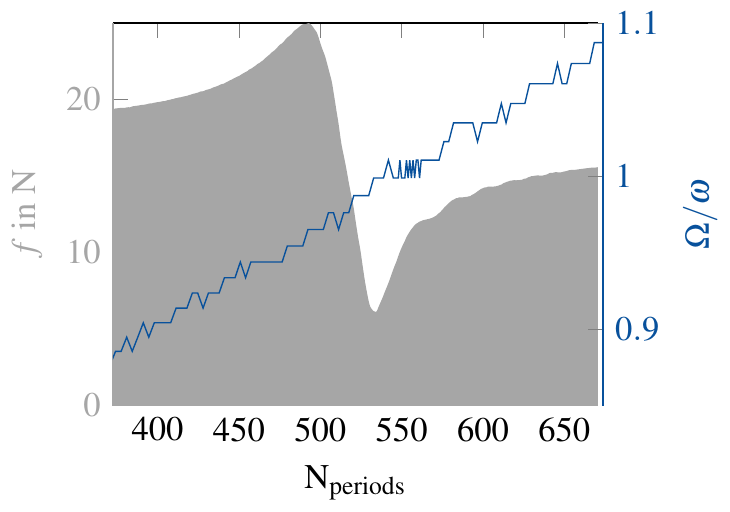}\caption{}
	\end{subfigure} 
	\begin{subfigure}{0.45\textwidth}
		\centering
		\includegraphics[width=0.8\textwidth]{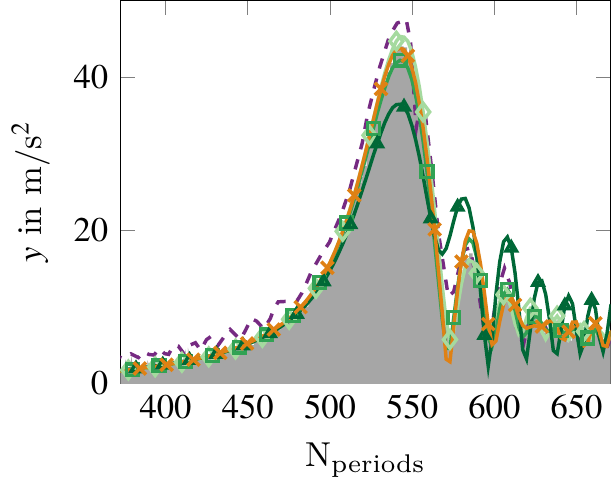}\caption{}
	\end{subfigure} 
	\begin{subfigure}{0.45\textwidth}
		\includegraphics[width=0.8\textwidth]{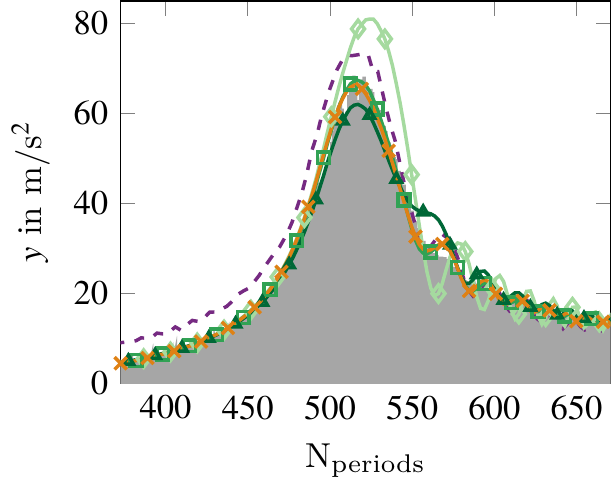}\caption{}
	\end{subfigure} 
	\begin{subfigure}{0.8\textwidth}
		\includegraphics[width=\textwidth]{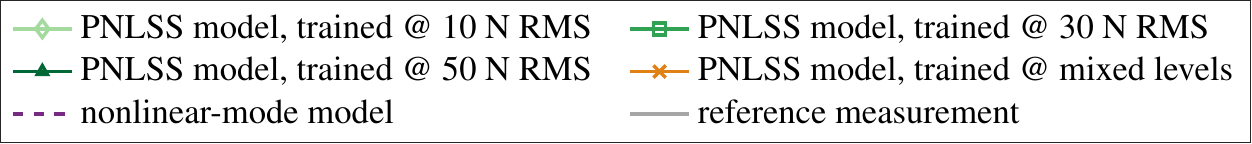}
	\end{subfigure} 
	\caption{(a)-(b) Measured excitation force and instantaneous excitation frequency. (c)-(d) Predicted response. The excitation level of sweep (a)/(c) is 8.1 N, and of sweep (b)/(d) 19.1 N.}\label{fig:BRB_sweep_syn}
\end{figure}

As previously, four mixed models are compared, now with respect to transient sweeps of the highest level (see \fref{BRB_transient_mixed}). All models predict the response with similar accuracy. It seems that \PNLSS models robustly predict transient sine sweeps for this specimen.

\begin{figure} % steady-state synthesis BRB
	\centering
	\includegraphics[width=\textwidth]{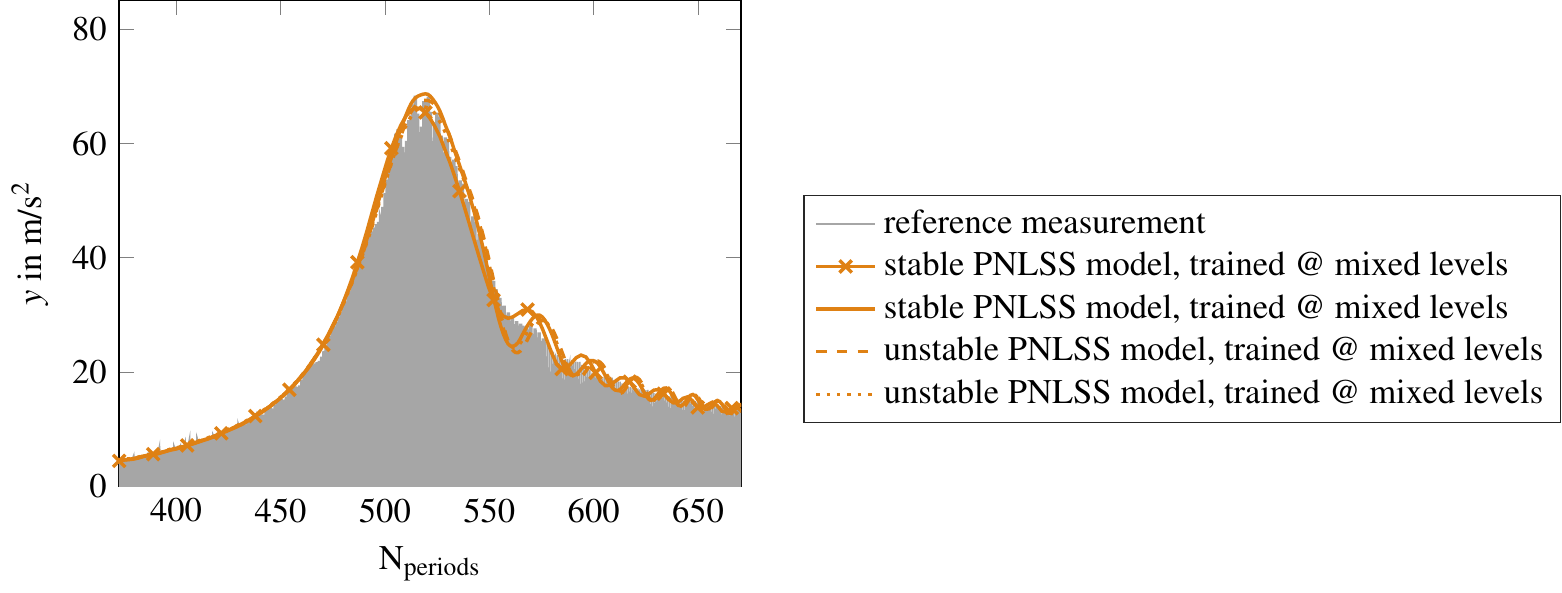}
	\renewcommand\figurename{Figure}
	\caption{Predicted response of transient sine sweeps \fref{BRB_sweep_syn} (b)/(d). The excitation level is 19.1 N. Comparison of four mixed models: one already shown in \fref{BRB_sweep_syn} (marked with crosses), another model that predicts bounded vibrations when tested with same-level multisine data and two models that predict unstable vibrations when tested with same-level multisine test data.}	\label{fig:BRB_transient_mixed}
\end{figure}

%%%%%%%%%%%%%%%%%%%%%%%%%%%%%%%%%%%%%%%%%%%%%%%%%%%%%%%%%%%%%%%%%%%%%%%%%%%%%%%%%%%%%%%%%%%%%
\section{Conclusions} \label{sec:conclusion}

In this paper, we study how predictive nonlinear data-driven models are with respect to vibrations close to an isolated resonance.
Two nonlinear system identification methods were chosen as representative examples, which employ fundamentally different training data for the identification:
With the first method, a single-nonlinear-mode model is identified based on periodic vibrations. With the second method, a \PNLSS model is identified using random broadband training data in this study. 
The identified models are then challenged to predict responses to different excitations, namely sine and sine sweep excitations. Steady-state sinusoidal excitation is close to the training data of the nonlinear-mode model (identified for periodic vibrations). Transient sweeps have only a single frequency at every time instant but cover a wider frequency band, where broadband training data for \PNLSS models might be beneficial. To fully assess the predictive capabilities of the models, they are additionally challenged to predict responses to excitation types \textit{different} from the training data.

The nonlinear-mode model is only valid around an isolated resonance and under essentially mono-harmonic excitation, which are clear model limitations. 
It performs well for predictions of steady-state responses, both for stiffness and damping nonlinearities. By design, there is a perfect match at phase resonance. In reality, repetition-variability and interpolation errors can lead to deviations.
Predicted transient vibrations had only moderate accuracy in this study; the reason for this was the high sensitivity to modal frequency errors if one ignores the physical mechanism behind the force drop typical for uncontrolled sweeps. 
Moreover, one must ensure that the time scales can indeed be separated, \ie that the frequency changes slowly compared to the oscillation.

Low-order polynomials approximate the real-life nonlinearities considered in this work only coarsely. Therefore, the identified \PNLSS models, and thus also their vibration prediction, depend largely on the training data. 
This dependence is the drawback of a black-box approach that does not require any a-priori knowledge on the system dynamics or type of nonlinearity. With increasing influence of nonlinear forces, the influence of modeling error and thus input-dependence increases concurrently.
Using training data of concatenated vibration levels seems to improve the model's validity range but still the model performs poorly in some cases. Transients sweeps are generally predicted with higher accuracy, especially for fast sweep rates. However, no \PNLSS model performs consistently well. All the more noteworthy, some identified models predict unstable behavior, even when tested with the same level as the training data. Yet, these models predict steady-state frequency responses or transient sine sweeps with reasonable accuracy in some cases. The input-dependence is more dominant for predictions of steady-state vibration than of transient sine sweeps. 
It is therefore still unclear how to define the valid regime of the \PNLSS model in order to prevent instability due to extrapolation. Moreover, the effect of extrapolation cannot be distinguished from model errors.

In principle, \PNLSS models could be improved by increasing the polynomial order. This, however, comes at the cost of a significant increase in computational time. 
In a follow-up study, we investigate if steady-state predictions of \PNLSS models achieve better agreement when the models are trained with steady-state signals, such as the training data of the nonlinear-mode model.

The experimental effort for identifying models with the two approaches depends on the required excitation signal. The modal approach requires a controller that needs to be tuned prior to the measurements. The measurement duration depends on this tuning. With systematic tuning, the measurement duration can be reduced.
The measurements for \PNLSS identification can be readily set up with commercial testing packages that offer multisine excitation.
In contrast, the identification of \PNLSS models requires nonlinear optimization, which comes with high computation effort and the difficulty of finding an appropriate initial guess. On the other side, the computational effort of identifying the nonlinear-mode model from the backbone measurements is low, where discrete Fourier transformation is the most involved step. In the present study, the measurements took a few minutes with either method for each specimen. The model identification of the nonlinear-mode models for both specimens and \PNLSS models for the beam with magnets took a few minutes, but it took a few hours to identify models for the \BRB. 

To summarize, we found two major limitations besides the above described capabilities: 
\begin{itemize}
	\item[-] The nonlinear-mode model is sensitive for uncontrolled sweeps if one ignores the physical mechanism behind the force drop.
	\item[-] \PNLSS models with only low-order polynomials are highly input-dependent, and may yield unbounded response even for input of the same level as training data. \PNLSS models can lead to qualitatively wrong predictions of steady-state vibrations for the nonlinearities considered in this study.
\end{itemize}
Only when bearing in mind these limitations, the methods can be applied successfully. We think that stating these limitations clearly assists the user in choosing a suited method for the individual application case and thus helps in mastering the challenging task of nonlinear system identification. Furthermore, stating these limitations hopefully triggers research to overcome these and thus further enhance the methods' capabilities.

\section*{Acknowledgment}

This work presents results of the \textit{Tribomechadynamics Research Camp}, formerly called \textit{Nonlinear Dynamics of Coupled Structures and Interfaces} summer school, an annual month long international research collaboration for graduate students and postdoctoral researchers. For more information, visit \texttt{http://tmd.rice.edu}. The work was funded by the Deutsche Forschungsgemeinschaft (DFG, German Research Foundation) - Project 402813361.

\appendix
\setcounter{figure}{0}
\setcounter{table}{0}

\section{PLL controller and parameters}\label{append:setup_parameters}

For the nonlinear modal analysis, the PLL controller sketched in \fref{PLL_scheme} was used.
For the reference measurement with stepped sines, the PLL was extended with a control loop for the force amplitude (see \fref{PLL_stepped_scheme}). The parameters for the controller are given in \tref{PLL_controller}. 

\begin{figure}
	\centering
	\begin{subfigure}{0.8\textwidth}
		\centering
		\includegraphics[width=\textwidth]{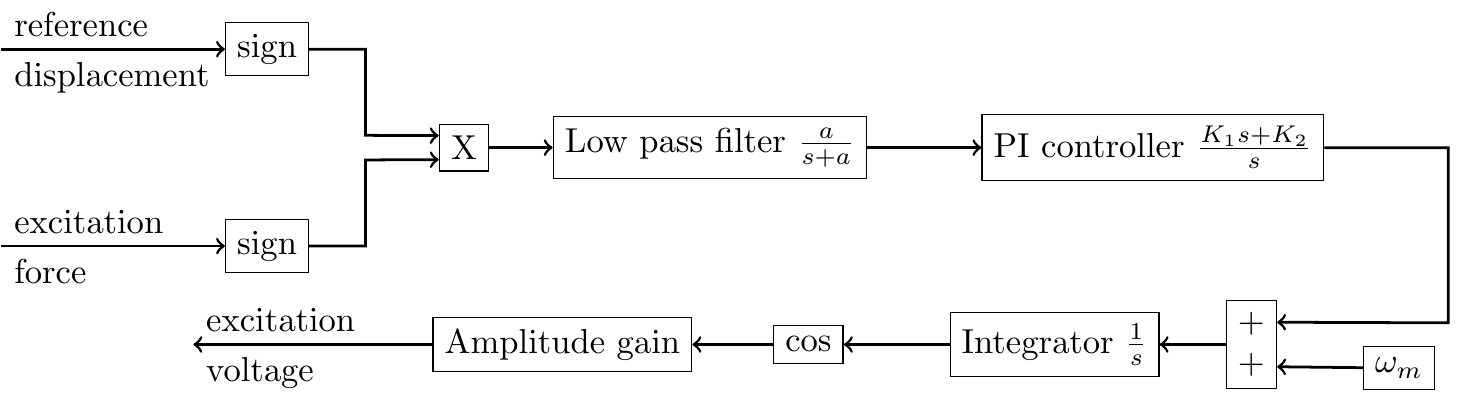}\caption{}\label{fig:PLL_scheme}
	\end{subfigure} 
	\begin{subfigure}{0.8\textwidth}
		\centering
		\includegraphics[width=\textwidth]{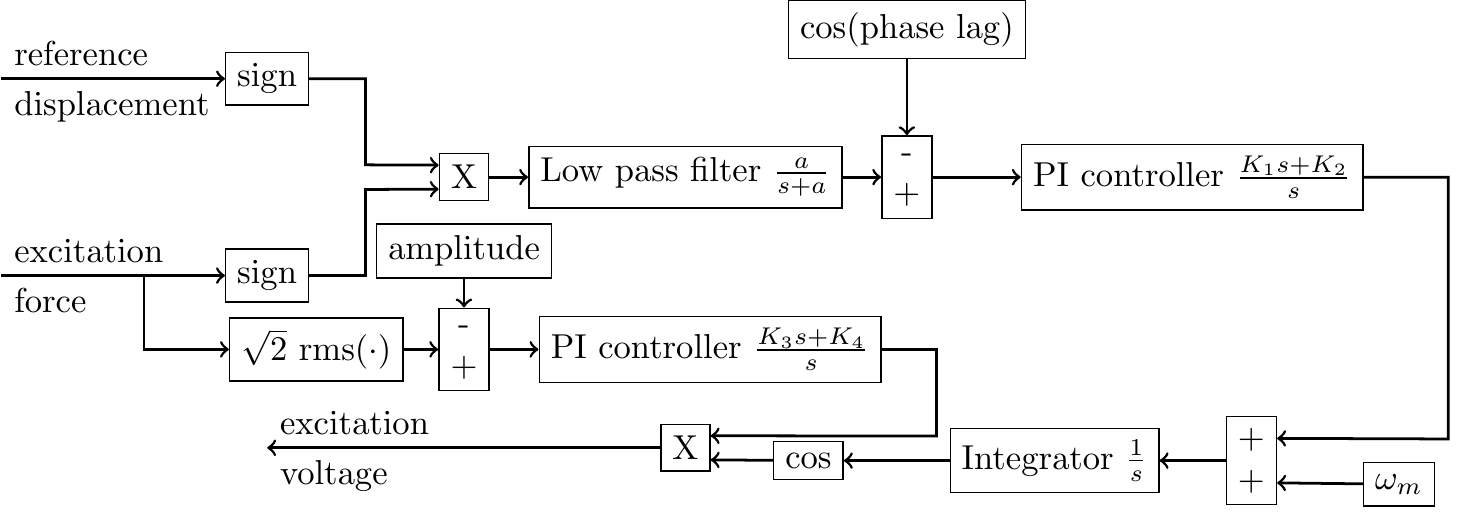}\caption{}\label{fig:PLL_stepped_scheme}
	\end{subfigure} 
	\caption{Scheme of PLL controller used for (a) nonlinear modal analysis and (b) stepped-sine measurements. The PLL in (b) is extended with a controller for the force amplitude.} 
\end{figure}

\begin{table} % parameters PLL
	\centering
	\begin{tabular}{l l l l l l l c}
		\toprule
		specimen & measurement & $a$ & $K_1$ & $K_2$ & $K_3$ & $K_4$ & $\omega_m$\\
		\midrule[0.8pt] 
		\multirow{2}{*}{beam with magnets}& stepped sine & $2 \pi$ & 2 & $2 \pi$ & 1.5 & 1 & $48 \pi$ rad/s\\
		& backbone & $2 \pi$ & 2 & $2 \pi$ & - & - & $48 \pi$ rad/s\\
		\midrule
		\multirow{2}{*}{BRB}& stepped sine & $2 \pi$ & 1 & $10\pi$ & 5 & 1 & $300 \pi$ rad/s\\
		& backbone & $2 \pi$ & 1 & $10\pi$ & - & - & $290 \pi$ rad/s\\
		\bottomrule		
	\end{tabular}
	\caption{Control parameters for measurements with PLL.}\label{tab:PLL_controller}
\end{table}

\bibliography{mssp_scheel_ref}

\end{document}